\documentclass[3p]{elsarticle}

\usepackage{lineno,hyperref}

\usepackage{csquotes}
\usepackage{todonotes}
\usepackage{bm}
\usepackage{amsmath}
\usepackage{subcaption}
\usepackage{booktabs}
\usepackage{graphicx}
\usepackage{grffile}
\usepackage{overpic}

\usepackage{color}
\newcommand{\rr}[1]{#1}

\newcommand{\rs}[1]{#1}

\newcommand{\ro}[1]{#1}
\newcommand{\rt}[1]{#1}

\hypersetup{
    colorlinks=true,
    linkcolor=blue,
    filecolor=magenta,      
    urlcolor=cyan,
}
 
\newcommand{\ten}[1]{{\bm #1}}
\renewcommand{\vec}[1]{{\bm #1}}

\journal{Communications in Nonlinear Science and Numerical Simulation}

\begin{document}

\begin{frontmatter}

\title{The effect of anisotropic viscosity on the nonlinear MHD kink instability}

\author[GlasgowAddress]{James Quinn\fnref{fn1}\corref{cor1}}
\ead{j.quinn.1@research.gla.ac.uk}
\cortext[cor1]{Corresponding author}
\author[GlasgowAddress]{David MacTaggart\fnref{fn2}}
\ead{David.MacTaggart@glasgow.ac.uk}
\author[GlasgowAddress]{Radostin \rs{D.} Simitev\fnref{fn3}}
\ead{Radostin.Simitev@glasgow.ac.uk}
\fntext[fn1]{\rs{\textit{ORCID:} \href{https://orcid.org/0000-0002-0268-7032}{orcid.org/0000-0002-0268-7032}}}
\fntext[fn2]{\rs{\textit{ORCID:} \href{https://orcid.org/0000-0003-2297-9312}{orcid.org/0000-0003-2297-9312}}}
\fntext[fn3]{\rs{\textit{ORCID:} \href{https://orcid.org/0000-0002-2207-5789}{orcid.org/0000-0002-2207-5789}}}

\address[GlasgowAddress]{School of Mathematics and Statistics, University of Glasgow, Glasgow G12 8QQ, UK}

\begin{abstract}
The kink instability of magnetohydrodynamics is believed to be fundamental to many aspects of the dynamic activity of the solar atmosphere, such as the initiation of flares and the heating of the solar corona. In this work, we investigate the importance of viscosity on the kink instability. In particular, we focus on two forms of viscosity; isotropic viscosity (independent of the magnetic field) and anisotropic viscosity (with a preferred direction following the magnetic field). Through the detailed analysis of magnetohydrodynamic simulations of the kink instability with both types of viscosity, we show that the form of viscosity has a significant effect on the nonlinear dynamics of the instability. The different viscosities allow for different flow and current structures to develop, thus affecting the behaviour of magnetic relaxation, the formation of secondary instabilities and the Ohmic and viscous heating produced. Our results have important consequences for the interpretation of solar observations of the kink instability.
\end{abstract}


\end{frontmatter}



\section{Introduction}

The solar corona is heated to millions of degrees, whereas the surface of the Sun beneath it is only heated to the order of thousands of degrees~\cite{mandriniMagneticFieldPlasma2000}. It is generally accepted that this heating is derived mainly from the extraction of energy from the stressed magnetic field in the solar atmosphere, although exactly how this process occurs is still an area of active research~\cite{klimchukSolvingCoronalHeating2006}. There has been much study of the direct transfer of magnetic energy to heat by Ohmic heating and the transfer through kinetic energy losses by viscous heating~\cite{klimchukSolvingCoronalHeating2006,browningMechanismsSolarCoronal1991}. The relative importance and efficacy of these processes is found to depend on the models of plasma viscosity employed in the analyses. The aim of this work is to understand the effect of isotropic and anisotropic viscosity on the kink instability, a key phenomenon in coronal plasma dynamics.

In many situations in the solar corona, the strength of viscosity
can be greater than that of resistivity by many orders of
magnitude, even when anomalous resistivity is considered. Due to this,
viscous heating can outperform Ohmic heating in certain coronal
situations, particularly those involving
reconnection~\cite{browningMechanismsSolarCoronal1991,
  craigViscousDissipation3D2013a,
  armstrongViscoResistiveDissipation2013,
  hollwegViscosityChewGoldbergerLowEquations1986}. These findings are
dependent on how viscosity in the solar atmosphere is modelled. For a
plasma in the presence of a strong magnetic field, viscosity is
anisotropic and momentum transfer occurs preferentially in the
direction of the magnetic
field~\cite{braginskiiTransportProcessesPlasma1965}. A general
description of anisotropic viscosity in the solar atmosphere is given
in~\cite{hollwegViscosityMagnetizedPlasma1985,
  hollwegViscosityChewGoldbergerLowEquations1986}. More recent
studies have demonstrated the importance of anisotropic viscosity for
heating in investigations of three-dimensional (3D) null
points~\cite{craigViscousDissipation3D2013a}, current sheet
merging~\cite{armstrongViscoResistiveDissipation2013} and flux
pile-up~\cite{litvinenkoViscousEnergyDissipation2005}. There is
further evidence of the importance of anisotropic viscosity in other
astrophysical applications, including the intracluster medium~\cite{zuhoneEffectAnisotropicViscosity2014, parrishEffectsAnisotropicViscosity2012a} and the solar wind~\cite{baleMagneticFluctuationPower2009}. In other solar applications, viscosity has a role to play in the damping of coronal instabilities~\cite{howsonEffectsResistivityViscosity2017} and waves~\cite{vranjesViscosityEffectsWaves2014, erdelyiResonantAbsorptionAlfven1995, rudermanSlowSurfaceWave2000}, though not all these cases have been fully explored using an anisotropic model of viscosity.

Implementing the full Braginskii viscosity
tensor~\cite{braginskiiTransportProcessesPlasma1965} into existing
magnetohydrodynamic (MHD) codes is not trivial. In the solar
atmosphere, there are regions where the magnetic field is weak or
vanishes (called null points) and at these locations the viscosity must
transition smoothly from strongly anisotropic (in the direction of the
magnetic field) to isotropic. Although the Braginskii tensor theoretically achieves
this smooth transition, in practice the implementation
of the full tensor in finite-difference MHD codes leads to numerical
errors in the viscosity at locations where the magnetic field strength
is very weak, due to a lack of sufficient resolution. Further
exploration of this numerical issue can be found
in~\cite{mactaggartBraginskiiMagnetohydrodynamicsArbitrary2017}. One
approach to solving this problem is to design a numerical scheme
specifically to treat the full Braginskii tensor. Another approach is
to devise a model that captures the main physics of the Braginskii
tensor suitable for the solar corona and can be implemented in
existing MHD codes~\cite{hollwegViscosityMagnetizedPlasma1985}. The
second option was taken up by MacTaggart, Vergori and
Quinn~\cite{mactaggartBraginskiiMagnetohydrodynamicsArbitrary2017},
who developed a phenomenological model of anisotropic viscosity in the
solar corona that captures the main physics for viscosity in the
corona, namely parallel viscosity in regions of strong field strength and isotropic viscosity in regions of very weak or zero field strength. In this paper, we will refer to this model of viscosity as the \textit{switching model}, for short. The model interpolates between isotropic and parallel viscosity based on how the local magnetic field strength affects the distribution of momentum transport. The interpolation itself can be adjusted, effectively changing the size of isotropic (i.e.\ weak field) regions, to compensate for resolution issues in simulations. In numerical tests of stressed null points, comparing the isotropic and anisotropic models showed that the use of isotropic viscosity overestimates the viscous heating \rt{by an order of magnitude~\cite{mactaggartBraginskiiMagnetohydrodynamicsArbitrary2017}.}

As a first step of investigating the effects of anisotropic viscosity on coronal dynamics, we focus on the kink instability~\cite{hoodKinkInstabilitySolar1979, hoodCoronalHeatingMagnetic2009}, believed to be a trigger for flares~\cite{srivastavaOBSERVATIONKINKINSTABILITY2010} and an important mechanism in the theory of coronal heating through nanoflares~\cite{browningHeatingCoronaNanoflares2008a}. The instability has also been studied using shock viscosity~\cite{hoodCoronalHeatingMagnetic2009,barefordShockHeatingNumerical2015} but a detailed investigation of the effects of Newtonian and Braginskii viscosity has not, to the best of our knowledge, been performed.
In particular, the main aim of this paper is to provide insight into the effect of the choice of viscosity model on the nonlinear dynamics and relaxation of a twisted coronal loop, where the kink instability converts magnetic energy to heat through Ohmic heating generated via current structures and through viscous heating generated via flow structures. We aim to give an estimate of how well viscous heating (using both isotropic and anisotropic models) performs when compared with Ohmic heating. This study extends previous work~\cite{hoodCoronalHeatingMagnetic2009} which also considers the kink instability in a zero-current loop (details given below). However, in contrast to~\cite{hoodCoronalHeatingMagnetic2009}, we consider only background resistivity and viscosity as the two mechanisms of heat generation (that is, we do not consider shock viscosity or anomalous resistivity). In performing this investigation, we also provide further validation of the switching model in a simpler topology to that used by MacTaggart et al.~\cite{mactaggartBraginskiiMagnetohydrodynamicsArbitrary2017}, in that there are no null points present in the field at any time.

The layout of the paper is as follows. The coronal loop model, MHD
equations and viscosity models are described in
Section~\ref{sec:model-setup}. \rr{Simulation details and quantities
  used in the analysis of the simulations are described in
  Section~\ref{sec:general-numerical-setup}.} Detailed numerical
results of a typical case of \rs{a kink instability} are presented in
Section~\ref{sec:results} with a particular focus on how the different
viscosity models affect \rs{its nonlinear evolution}. The results of
the typical case are \rs{confirmed and} generalised \rs{by a parameter
  study} in Section~\ref{sec:results2} \rs{where the dependences of the
Ohmic and the viscous heating on the resistivity and the dynamic
viscosity are presented.} Our conclusions are summarised in Section~\ref{sec:conclusions}.

\section{Models of kink instability and viscosity}
\label{sec:model-setup}

As a model of an idealised coronal loop, we consider a
twisted flux tube in a Cartesian box of dimension $[-2\text{
    Mm},2\text{ Mm}] \times [-2\text{ Mm},2\text{ Mm}] \times
   [-10\text{ Mm},10\text{ Mm}]$ in the $x$, $y$ and $z$-directions,
   respectively. The state of the plasma is typical of the corona,
   with density $\rho$ initially $1.67\times 10^{-12} \text{
     kgm}^{-3}$, and with plasma pressure $p$ such that the
   temperature of the plasma $T$ is initially $2\times10^{4} \text{
     K}$ everywhere in the computational domain. The magnetic
   field $\vec{B}$ is constructed so that it is initially force-free
   and with zero axial current, line-tied at the boundaries, and
   twisted such that it is linearly unstable to the ideal kink
   instability. This configuration allows us to compare directly to
   previous studies that use identical magnetic field
   configurations~\cite{hoodCoronalHeatingMagnetic2009,barefordShockHeatingNumerical2015,bothaOBSERVATIONALSIGNATURESCORONAL2012}. The
   field outside the flux tube is straight and has a strength of
   $5\times10^{-3} \text{ T}$. Given this temperature and magnetic
   field strength, the plasma beta is initially $\beta \approx
   10^{-5}$, a value realistic for the corona. \rs{The evolution} of
   this flux tube is \rs{governed by the nonlinear} MHD equations.

\subsection{Model equations}

Non-dimensionalised, the MHD equations take the form
\begin{gather}
\label{eq:mhda}
\frac{D\rho}{Dt} = - \rho \vec{\nabla} \cdot \vec{u},\\
\rho\frac{D\vec{u}}{Dt} = -\vec{\nabla} p + \vec{\jmath} \times \vec{B} + \vec{\nabla} \cdot \ten{\sigma},\\
\frac{D\vec{B}}{Dt} = (\vec{B} \cdot \vec{\nabla})\vec{u} - (\vec{\nabla} \cdot \vec{u})\vec{B} + \eta \nabla^2 \vec{B},\\
\rho\frac{D\varepsilon}{Dt} = -p \vec{\nabla} \cdot \vec{u} + {Q}_{\nu} + {Q}_{\eta},
\label{eq:energy}
\end{gather}
where \rr{$\vec{u}$ is the plasma velocity,} $\vec{\jmath} = \nabla
\times \vec{B}$ is the current density, $\ten{\sigma}$ is the viscous
stress tensor, $\eta = 1/S$ is the normalised resistivity (equivalent
to the inverse of the Lundquist number $S$), and use has been made of
the material derivative, $D/Dt = \partial/\partial t + (\vec{u} \cdot
\vec{\nabla})$. \rs{The internal energy density is given by the equation of state for an ideal gas}
\begin{equation}
\varepsilon = \frac{p}{\rho(\gamma - 1)},
\end{equation}
where the specific heat ratio is given by $\gamma = 5/3$. \rs{The
  terms ${Q}_{\nu} = \ten{\sigma} : \vec{\nabla}\vec{u}$ and
  ${Q}_{\eta} = \eta | \vec{\jmath} |^2$ model Ohmic heating and viscous heating, respectively.}

Using the nondimensionalisation scheme found
in~\cite{arberStaggeredGridLagrangian2001}, reference values for the
magnetic field $B_0$, length $L_0$ and density $\rho_0$ are chosen to
align with values typical for a coronal loop. The problem domain is
non-dimensionalised to $[-2,2] \times [-2,2] \times [-10,10]$ in the
$x$, $y$ and $z$-directions, respectively. Velocity and time are
non-dimensionalised using the Alfv\'en speed $u_A = B_0 / \sqrt{\rho_0
  \mu_0}$ and Alfv\'en crossing time $t_A = L_0/u_A$,
respectively. Temperature is non-dimensionalised via $T_0 = u_A^2
\bar{m} / k_B$, where $k_B$ is the Boltzmann constant and $\bar{m}$ is
the average mass of ions, here taken to be $\bar{m} = 1.2m_p$ (a mass
typical for \rr{the solar corona}) where $m_p$ is the proton mass. The
reference values of other variables are derived from these
\rr{quantities} and \rs{are listed} in Table~\ref{tab:reference-values}. Dimensional quantities can be recovered by multiplying the non-dimensional variables by their respective reference value (e.g. $\vec{B}_{\dim} = B_0 \vec{B}$). All further reference to variables will be to their non-dimensionalised values, unless stated otherwise.

\begin{table}[t]
\centering
\begin{tabular}{ccc|ccc}
$B_0$ & $L_0$ & $\rho_0$ & $u_A = B_0 / \sqrt{\rho_0 \mu_0}$ & $t_A = L_0/u_A$ & $T_0$ \\ \midrule
$5 \times 10^{-3} \ \text{T}$ & $1\ \text{Mm}$ & $1.67 \times 10^{-12} \ \text{kgm}^{-3}$ & $3.45\ \text{Mms}^{-1}$ & $0.29\ \text{s}$ & $1.73 \times 10^{9}K$\\
\end{tabular}
\caption{Reference values for the magnetic field, length, density, and
  temperature. \rs{These are} used to non-dimensionalise the MHD
  equations \rs{\eqref{eq:mhda}--\eqref{eq:energy} and} to calculate the reference values for velocity, time and temperature.}
\label{tab:reference-values}
\end{table}

\rs{We consider a force-free magnetic field for which the Lorentz
  force is zero so that $(\nabla \times
  \vec{B})\times \vec{B} = 0$.}
To satisfy this condition we choose, \rs{in cylindrical coordinates
  $(r,\theta,z)$, a magnetic field of the form} $\vec{B} = \alpha(r) \nabla \times \vec{B}$, where
$\alpha(r)$ is a function of a particular form that ensures the total
axial current is zero. Aligning with previous work by Hood et
al.~\cite{hoodCoronalHeatingMagnetic2009}, the smooth $\alpha(r)$
profile given as Case 3 in~\cite{hoodCoronalHeatingMagnetic2009} is
used. Using this profile, the equilibrium magnetic field $\vec{B}$ is
written  as
\begin{equation}
\begin{aligned}
  \label{eq:field-profile-r-lt-1}
  B_{\theta} &= \lambda r {(1 - r^2)}^3,\\
  B_z &= \sqrt{1 - \frac{\lambda^2}{7} + \frac{\lambda^2}{7}{(1 - r^2)}^7 - \lambda^2 r^2 {(1-r^2)}^6},\\
  \alpha(r) &= \frac{2 \lambda {(1-r^2)}^2 {(1-4r^2)}}{B_z},
\end{aligned}
\end{equation}
for $r \leq 1$ and
\begin{equation}
\begin{aligned}
  \label{eq:field-profile-r-gt-1}
  B_{\theta} &= 0 \\
  B_z &= \sqrt{1 - \frac{\lambda^2}{7}}\\
  \alpha(r) &= 0 ,
\end{aligned}
\end{equation}
for $r > 1$, where $\lambda$ is a parameter measuring the twist in the tube. The radial field throughout the domain is set to $B_r = 0$. As is done in~\cite{hoodCoronalHeatingMagnetic2009}, we set $\lambda = 1.8$ to ensure the tube is unstable to the ideal kink instability. The equilibrium velocity for this magnetic field configuration is $\vec{u} = \vec{0}$.

\begin{figure}[t]
  \centering
  \begin{subfigure}[b]{0.48\textwidth}
  \begin{center}
    \begin{overpic}[width=\textwidth]{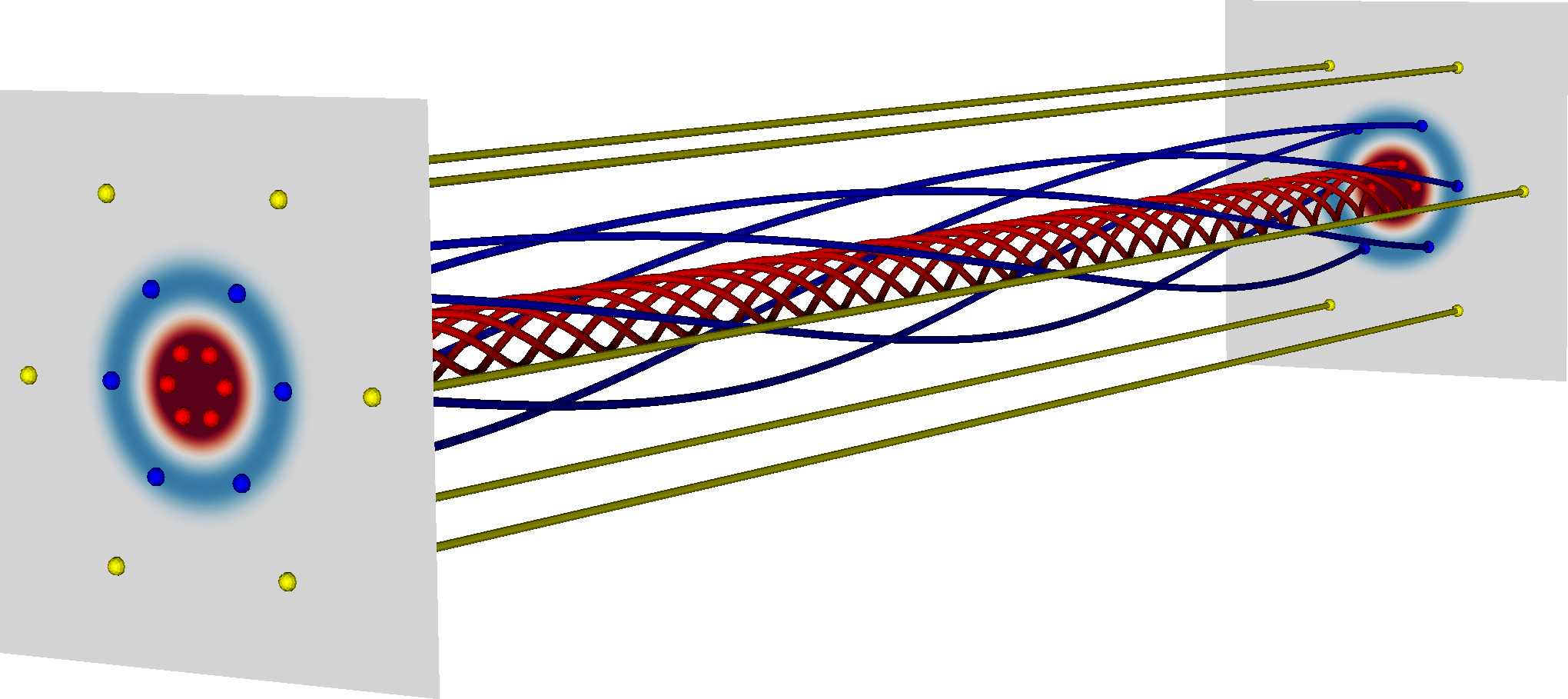}
      \put (50,5) {\small\textbf{(a)}}
    \end{overpic}
  \end{center}
  \end{subfigure}
  \begin{subfigure}[b]{0.48\textwidth}
  \begin{center}
    \begin{overpic}[width=\textwidth]{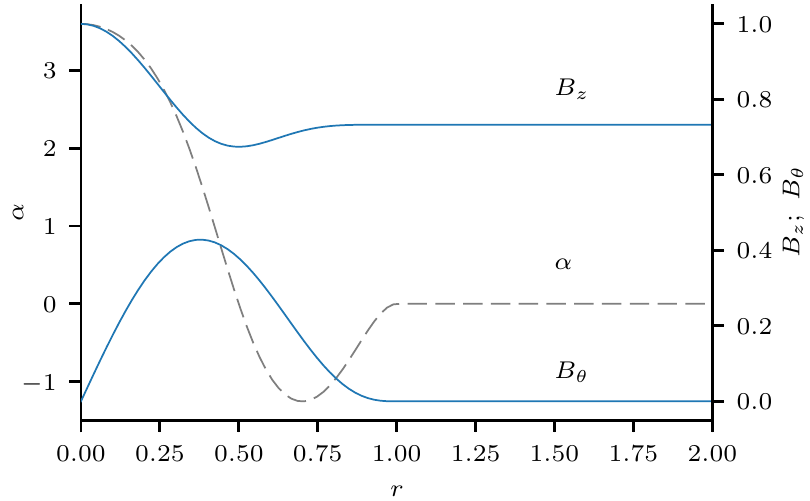}
      \put (47,60) {\small\textbf{(b)}}
    \end{overpic}
  \end{center}
  \end{subfigure}
  \caption{\textit{The initial field configuration.} In \textbf{(a)} field lines are plotted corresponding to inner (red), outer (blue) and straight (yellow) regions of twist, with slices of $\alpha(r)$ shown at the footpoints. In the slices, red corresponds to $\alpha(r) > 0$, blue to $\alpha(r) < 0$ and white to $\alpha(r) = 0$. In \textbf{(b)} the profiles of $\alpha(r)$ and the field components $B_z$ and $B_{\theta}$ across the flux tube are plotted.}
\label{fig:field_configuration}
\end{figure}

The form of $\alpha(r)$ in equations~\eqref{eq:field-profile-r-lt-1} and~\eqref{eq:field-profile-r-gt-1} splits the profile of the flux tube into three twist regions, the inner region of positive twist ($r\le0.5$), the outer region of negative twist ($0.5<r<1$) and the straight-field region \rs{of} zero twist ($r\ge1$) as shown in Figures~\ref{fig:field_configuration}(a) and (b). These figures also illustrate the equilibrium field. Since the inner region is more tightly twisted, this field configuration results in only the inner region becoming unstable to the kink instability, rather than the global instability seen in non-zero-current loops~\cite{hoodKinkInstabilitySolar1979}. The regions of twist are used later to define a measure of reconnection.

Although we prescribe an initial temperature \rr{of} $T=2\times10^{4} \text{ K}$, the equations simulated by the code are written using internal energy, thus we convert the temperature to internal energy using the non-dimensional relation $\varepsilon = T/(1-\gamma)$. Hence, the initial non-dimensionalised density and internal energy are \rs{uniformly} given by
\begin{equation}
  \rho = 1,\quad \varepsilon = 8.66 \times 10^{-4},
\end{equation}
and have been non-dimensionalised using the reference values found in Table~\ref{tab:reference-values}. The initial magnetic field and velocity are set to their equilibrium states, discussed above, with the addition of a small perturbation.

In order to \rt{make a meaningful comparison of our results} with those of~\cite{hoodCoronalHeatingMagnetic2009}, we use identical initial magnetic field and velocity perturbations, calculated via a linear stability analysis (in ideal MHD) applied to a similar flux tube that uses a constant, piecewise profile for $\alpha(r)$~\cite{vanderlindenCompleteCoronalLoop1999,browningSolarCoronalHeating2003,browningHeatingCoronaNanoflares2008a}.

At the boundaries, we satisfy the line-tied condition on the magnetic field by ensuring the field is constant and equal to its initial values given by equations~\eqref{eq:field-profile-r-lt-1} and~\eqref{eq:field-profile-r-gt-1}. Similarly, on the boundaries we ensure that the density, internal energy and velocity $\vec{u}$ are constant and equal to their initial values. To close the system, the fluxes of all variables through each of the boundaries are set to zero. That is, on the $x$-boundary,
\begin{equation}
  \frac{\partial \vec{B}}{\partial x} = \frac{\partial \vec{u}}{\partial x} = \vec{0}; \quad \frac{\partial \rho}{\partial x} = \frac{\partial \varepsilon}{\partial x} = 0 \quad \text{for } x=\pm 2,
\end{equation}
and similarly, the $y$ and $z$ derivatives are zero on the $y=\pm2$ and $z=\pm10$ boundaries, respectively.

\subsection{Anisotropic viscosity}

Isotropic (Newtonian) viscosity is the most commonly used viscosity model for applications to the solar corona. Its viscous stress tensor is directly proportional to the rate-of-strain of a flow through the dynamic viscosity of the fluid, $\nu$,
\begin{equation}
\ten{\sigma}_{\text{iso}} = \nu \ten{W},
\label{eq:isotropic_stress_tensor}
\end{equation}
where $\ten{W} = \vec{\nabla \vec{u}} + {(\vec{\nabla \vec{u}})}^{\text{T}} - \frac{2}{3}(\vec{\nabla} \cdot \vec{u}) \ten{I}$ is the traceless rate-of-strain tensor and $\ten{I}$ is the identity tensor.
\rs{In the solar corona,} apart from regions of very weak field strength, viscosity is dominated by the strong field limit of the full Braginskii tensor~\cite{hollwegViscosityChewGoldbergerLowEquations1986},
\begin{equation}
\ten{\sigma}_{\text{strong}} = \frac{3}{2} \nu (\ten{W}\vec{b}) \cdot \vec{b} \left( \vec{b} \otimes \vec{b} - \frac{1}{3}\ten{I}\right),
\label{eq:strong_braginskii_stress_tensor}
\end{equation}
where $\vec{b} = \vec{B}/|\vec{B}|$ is a unit vector in the direction of the magnetic field. \ro{The strong field tensor can be considered made up of two parts: the scalar strength of viscosity, given by the multiplication of the viscosity coefficient $\nu$ and the interaction of the strain rate tensor with the field $(\ten{W} \vec{b}) \cdot \vec{b}$; and the direction of action, involving \emph{only} the direction of $\vec{b}$.
To illustrate how velocity gradients affect this tensor, consider a magnetic field  where the frame of reference is such that the field is oriented in the $z$-direction, that is $\vec{b} = {(0,0,1)}^T$. In this frame,
\begin{equation}
  (\ten{W} \vec{b}) \cdot \vec{b} = \ten{W}_{33} = \frac{2}{3}\left(3\frac{\partial u_z}{\partial z} - \nabla \cdot \vec{u} \right),
\end{equation}
and the full tensor in equation~\eqref{eq:strong_braginskii_stress_tensor} can be written, in matrix form, as
\begin{equation}
[\ten{\sigma}_{\text{strong}}] = \frac{\nu}{3}\left(3\frac{\partial u_z}{\partial z} - \nabla \cdot \vec{u} \right)
  \begin{bmatrix}
  -1 & 0 & 0 \\
  0 & -1 & 0 \\
  0 & 0 & 2
  \end{bmatrix}.
  \label{eq:parallel_tensor}
\end{equation}
Consider a velocity only in the $z$-direction, $u_z$, exhibiting a gradient in the same direction, $\partial u_z / \partial z$. In this case, equations~\eqref{eq:isotropic_stress_tensor} and~\eqref{eq:strong_braginskii_stress_tensor} are equivalent, hence the anisotropic viscosity acts identically to isotropic viscosity in the direction of the field. In contrast, if that same velocity $u_z$ exhibits only a gradient perpendicular to the field, say $\partial u_z / \partial x$, there would be no strong field viscosity, since the shear gradient does not enter into expression~\eqref{eq:parallel_tensor}, and the flow would be undamped. Similarly, if the flow consisted of only perpendicular velocities with gradients directed along the field, the viscosity would vanish and these flows would remain undamped. Indeed, a flow exhibiting only shear velocity gradients would remain undamped. For a further exploration of the behaviour of this tensor, we refer to the more comprehensive discussion of the terms of the full Braginksii tensor in~\cite{braginskiiTransportProcessesPlasma1965} and to the physical interpretation found in~\cite{hollwegViscosityMagnetizedPlasma1985}.
}
\rs{As mentioned previously,} the switching model of MacTaggart et al.~\cite{mactaggartBraginskiiMagnetohydrodynamicsArbitrary2017} interpolates between equations~\eqref{eq:isotropic_stress_tensor} and~\eqref{eq:strong_braginskii_stress_tensor},
\begin{equation}
\ten{\sigma}_{\text{aniso}} = [1 - s^2(|\vec{B}|)]\ten{\sigma}_{\text{iso}} + s^2(|\vec{B}|)\ten{\sigma}_{\text{strong}}.
\label{eq:switching_stress_tensor}
\end{equation}
The interpolation function $s(|\vec{B}|)$ is derived in~\cite{mactaggartBraginskiiMagnetohydrodynamicsArbitrary2017} by considering how the distribution of momentum transport depends on the magnetic field strength. The interpolating function is given by
\begin{equation}
s(|\vec{B}|) = \frac{3 \exp[2a]}{2\sqrt{2\pi a} \text{erfi}[\sqrt{2a}]} - \frac{1}{2}\left[ 1 + \frac{3}{4a} \right],
\label{eq:s-function}
\end{equation}
where $a(|\vec{B}|)$ is a constitutive function, chosen here to take its simplest, non-trivial form $a(|\vec{B}|) = a_0 |\vec{B}|^2$. In practial terms, the parameter $a_0$ varies the size of the region where isotropic viscosity dominates in the numerical simulations. To align with earlier work in~\cite{mactaggartBraginskiiMagnetohydrodynamicsArbitrary2017}, we choose $a_0 = 150$. \ro{The application of the switching model in simulations of a stressed null point in~\cite{mactaggartBraginskiiMagnetohydrodynamicsArbitrary2017} provides a fuller exploration of the isotropic feature of the model. Despite the switching itself being relatively unimportant here due to viscosity remaining fully anisotropic nearly everywhere, we still choose to use the switching model for two reasons: consistency with previous work~\cite{mactaggartBraginskiiMagnetohydrodynamicsArbitrary2017}, and as a tool to investigate the effect of prescribing anisotropy, which we will discuss in more detail later.}

The switching viscosity is not only a physically realistic model for
anisotropic viscosity in the corona, but can \rs{also} be used to investigate, more deeply, the differences between isotropic and anisotropic viscosity. This is done by artificially setting the result of $s$ in the code to fix the amount of anisotropy present. We will consider the balance between isotropic and anisotropic viscosity in more detail later.


\section{Numerical setup and tools of analysis}
\label{sec:general-numerical-setup}

Here we describe the \rs{computational} code used to \rs{solve numerically} the
governing equations and we define the \rs{diagnostic} tools used to
analyse the \rs{simulation} output \rs{results}.

\subsection{Numerical setup}

We solve the \rs{governing} MHD \rs{equations~\eqref{eq:mhda}---\eqref{eq:energy} numerically} using the Lare3d
code~\cite{arberStaggeredGridLagrangian2001}, which is widely used in
the solar physics research community. Lare3d is based on a
Lagrangian-Remap scheme with artificial viscosity and flux-limiters as
shock-capturing devices. Since we are investigating the role of
viscosity itself, the artificial viscosity (or shock viscosity) has been disabled. \rs{In
order to compare features of our simulations with the results of Hood
et al.~\cite{hoodCoronalHeatingMagnetic2009} we have performed} tests
using shock viscosity, instead of either the switching or isotropic models. Using the default shock viscosity parameters present in the code, the behaviour closely mirrors that of isotropic viscosity with $\nu\approx 5\times10^{-4}$. When both switching and shock viscosity are enabled, the shock viscosity dominates and, again, the behaviour mirrors that of isotropic viscosity.

The simulations were run at a resolution of $350 \times 350 \times 700$, with the exception of the parameter studies, which were run at a slightly higher resolution of $400 \times 400 \times 800$. Since the switching viscosity only acts parallel to the magnetic field, in perpendicular directions numerical diffusion dominates. By running several simulations at resolutions of $250 \times 250 \times 500$ up to $500 \times 500 \times 1000$, it was found that the effect of resolution was negligibly small until around $t=150$, well after the nonlinear phase of the instability. After this time there are some quantitative differences in outputs for different resolutions. However, the qualitative behaviour, which we describe later, does not strongly depend on the resolution.

We use an estimate of $\tilde{\nu} = \tilde{\eta} = \tilde{u}_x L_x/N_x^2$ for the numerical diffusion present in the simulations due to the finite difference scheme employed in Lare3d. Taking a typical velocity of $\tilde{u}_x = 1$, i.e.\ the Alfv\'en velocity; $N_x = 350$ as the number of grid-points in the $x$-direction; and $L_x = 4$ as the length in the $x$ direction, we estimate the numerical diffusion coefficient to be $\tilde{\nu} = \tilde{\eta} \approx 10^{-5}$. This provides a theoretical lower bound on simulating a physical viscosity or resistivity. In practice, however, we find setting the physical resistivity lower than $\eta \approx 5\times10^{-5}$ results in behaviour that \rr{does not} converge with increasing resolution. This gives a practical lower bound for diffusion coefficients of $\tilde{\nu} = \tilde{\eta} \approx 5 \times 10^{-5}$. Thus, all results presented use physical diffusion coefficients (either viscosity or resistivity) greater than this lower bound.

\subsection{Connectivity}

As part of the analysis in Section~\ref{sec:results}, we present a
practical measure of magnetic reconnection --- the mean change in
field line connectivity $\Delta\Phi_c$. We determine the connectivity
$\Phi_c$ for field lines by analysing the start and end points of a
sample of magnetic field lines. Any field line that begins at one
location in one of the footpoints will map to a corresponding point in
the opposite footpoint. Following field lines from their starting
point at $z=-10$ to their end point at $z=10$, we label lines
depending on the twist regions in which they start and end. Initially,
the field lines within each distinct region map one-to-one to the same
region. As the field reconnects radially during the instability, field lines
begin to start and end in different twist zones. By tracking the
number of field lines that have changed twist zones within a \rs{set}
time \rs{period} we get a practical estimate for the rate of
reconnection, \rs{as well as} a visual guide of where that reconnection is
occurring. \ro{It should be noted that this measure of reconnection does not take into account azimuthal reconnection (that is reconnection within the same twist zone). As such, it is only a partial measure of reconnection.}

In practice, magnetic field \rs{output is saved} from the code at
intervals of $\Delta t = 5$. We use the visualisation tool
Mayavi~\cite{ramachandran2011mayavi} to compute the magnetic field
lines over a grid of starting points $(x_i, y_j)$ at a given time $n
\Delta t$, where $n$ indexes the output files. This process gives a
connectivity map $\Phi_c^{(n)}(x_i, y_j)$ across the profile of the
flux tube. \rs{The mean} difference in connectivity $\Delta
\Phi_c^{(n)}$ at time $n\Delta t$ \rs{is found} by subtracting one connectivity map from the previous and then taking the mean across all points $(x_i, y_j)$,
\begin{equation}
  \Delta \Phi_c^{(n)} = \frac{1}{N_x N_y} \sum_{i=1}^{N_x} \sum_{j=1}^{N_y} (\Phi_c^{(n)}(x_i, y_j) - \Phi_c^{(n-1)}(x_i, y_j)).
\end{equation}

\subsection{Parallel electric field}

Another useful measure of magnetic reconnection is the maximum value of the integral of the electric field parallel to the magnetic field $E_{\parallel} = \eta {(\vec{\jmath} \cdot \vec{B})}/|\vec{B}|$ along a magnetic field line~\cite{galsgaardSteadyStateReconnection2011,priestNatureThreedimensionalMagnetic2003,schindlerGeneralMagneticReconnection1988},
\begin{equation}
  \Phi = \int_{C} \eta \frac{(\vec{\jmath} \cdot \vec{B})}{|\vec{B}|}\ {\rm d}l,
\end{equation}
where $C$ is a magnetic field line with start and end points within the footpoints at $z\pm10$.

\rs{Similarly to the calculation of connectivity, we employ} Mayavi to
compute magnetic field lines \rs{using a grid of field line starting
  points $(x_i, y_j)$ at a given time}. We track the local value of
the \rs{modulus} of the parallel electric field, $|E_{\parallel}| =
|\eta \vec{\jmath} \cdot \vec{B}|$ along the magnetic field lines. The
parallel electric field is then summed along each of the field lines
to give a \rs{distribution $\Phi(x_i, y_j)$} across the profile of the field. The maximum of this distribution gives a measure of the reconnection rate.

It can be argued that the reconnection rate calculated by taking the global maximum is only the rate for one region of magnetic diffusion, and the nonlinear phase of the kink instability creates multiple diffusion regions in its development. One way to calculate the reconnection rate for each region is via the algorithm described in~\cite{pontinDynamicsBraidedCoronal2011}, which dissects the distribution $\Phi(x_i, y_j)$ into separate regions before finding the maxima corresponding to the reconnection rate per diffusion region. In practice, we find the current structures created by the kink instability to be simple enough that this extended analysis is unnecessary.

\subsubsection{Other observables}

\rs{To further characterise} our results we use of the volume-integrated parallel and perpendicular kinetic energies,
\begin{equation}
  \label{eq:kinetic_energies}
  \text{KE}_{\parallel} = \frac{1}{2} \int_V \rr{\rho}\frac{(\vec{u}\cdot\vec{B})^2}{|\vec{B}|^2}\ \text{d}V; \quad
  \text{KE}_{\perp} = \frac{1}{2} \int_V \rr{\rho}|\vec{u}|^2\ \text{d}V - \text{KE}_{\parallel},
\end{equation}
\rr{the} magnetic energy,
\begin{equation}
  \label{eq:magnetic_energy}
   \text{ME} = \rr{\frac12}\int_V |\vec{B}|^2\ \text{d}V,
\end{equation}
and the total Ohmic heating generated by time $T$,
\begin{equation}
  \label{eq:ohmic_heating}
  Q_{\eta} = \eta \int_0^{T} \int_V |\vec{\jmath}|^2\ \text{d}V \text{d}t.
\end{equation}
The time and volume-integrated viscous heating rate can be written
\rs{in the form}
\begin{equation}
  \label{eq:iso_viscous_heating}
  Q_{\nu}^{iso} = \frac{\nu}{2} \int_0^T \int_V
  \text{tr}(\ten{W}^2)\  \text{d}V \text{d}t,
\end{equation}
for the isotropic viscous stress
tensor~\eqref{eq:isotropic_stress_tensor} and \rs{in the form}
\begin{equation}
  \label{eq:aniso_viscous_heating}
  Q_{\nu}^{aniso} = \nu \int_0^T \int_V \left[ (1-s^2(|\vec{B}|)\frac{1}{2}\text{tr}(\ten{W}^2) + s^2(|\vec{B}|)\frac{3 }{4} ((\ten{W} \vec{b}) \cdot \vec{b})^2\ \right] \text{d}V \text{d}t,
\end{equation}
for the switching viscous stress tensor \eqref{eq:switching_stress_tensor}, \rs{respectively.}

\section{Nonlinear evolution of a typical case}
\label{sec:results}

\rs{In this section, we} present results from simulations with a
specific choice of viscosity and resistivity. This provides an
opportunity to analyse, in detail, the onset and evolution of the kink
instability in a typical case. Parameter studies illustrating that
\rs{the observed dynamics is} typical are presented \rs{further below}
in Section~\ref{sec:results2}.
\rs{We compare} the results from simulations of two
cases. \rs{Isotropic viscosity is used in the first case and switching
  viscosity is used in the second case}. The choice of viscosity
tensor is the only difference between the two cases. \rs{In this
  section,} we use the diffusion parameters $\nu = 10^{-4},\ \eta =
5\times 10^{-4.5}$, both small but suitably above the threshold of
numerical diffusion discussed in
Section~\ref{sec:general-numerical-setup}. \ro{The chosen value of $\nu$ is well within the range of typical values found in the real corona, that is, in our nondimensionalisation, between $10^{-8}$ and $10^{-3}$~\cite{rudermanSlowSurfaceWave2000}.} All other parameters are
identical in both cases and are kept fixed to the values specified in
Section~\ref{sec:general-numerical-setup}. \ro{Due to the strength of the field and lack of null points, it is measured that $s=1$ throughout the entire domain, thus the switching model reverts to the strong field approximation of the Braginskii tensor~\eqref{eq:strong_braginskii_stress_tensor}.}

\subsection{Linear phase}

\begin{figure}[t]
  \begin{subfigure}[b]{0.48\textwidth}
\hspace*{-5mm}
\vspace{5mm} 
   \begin{overpic}[width=\textwidth]{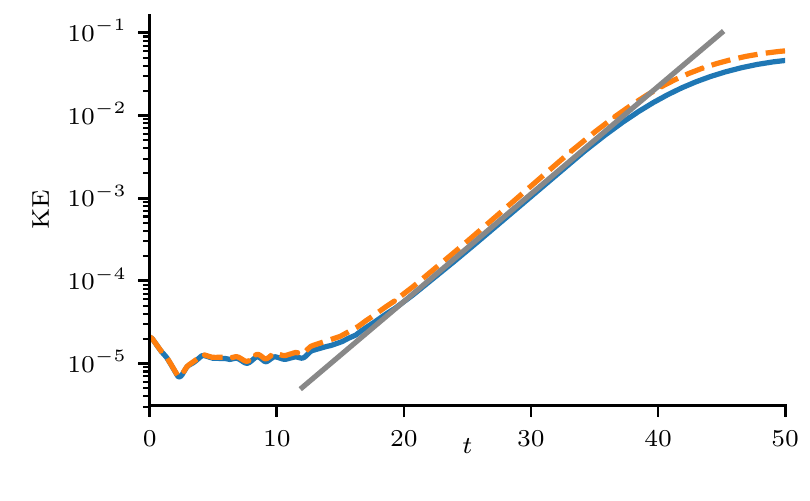}
      \put (50,60) {\small\textbf{(a)}}
    \end{overpic}
  \end{subfigure}
\hfill
 \begin{subfigure}[b]{0.48\textwidth}
    \begin{overpic}[width=\textwidth]{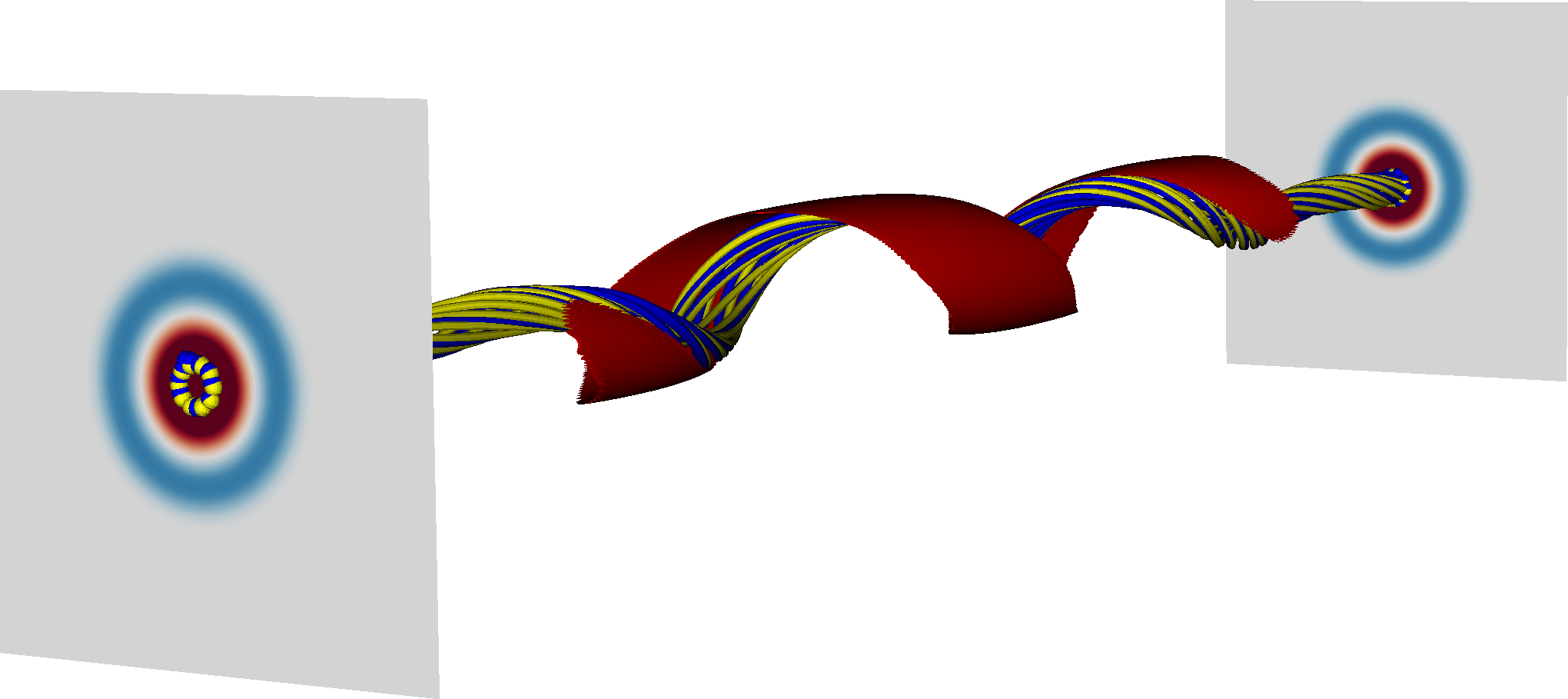}
      \put (50,40) {\small\textbf{(b)}}
    \end{overpic}
    \begin{overpic}[width=\textwidth]{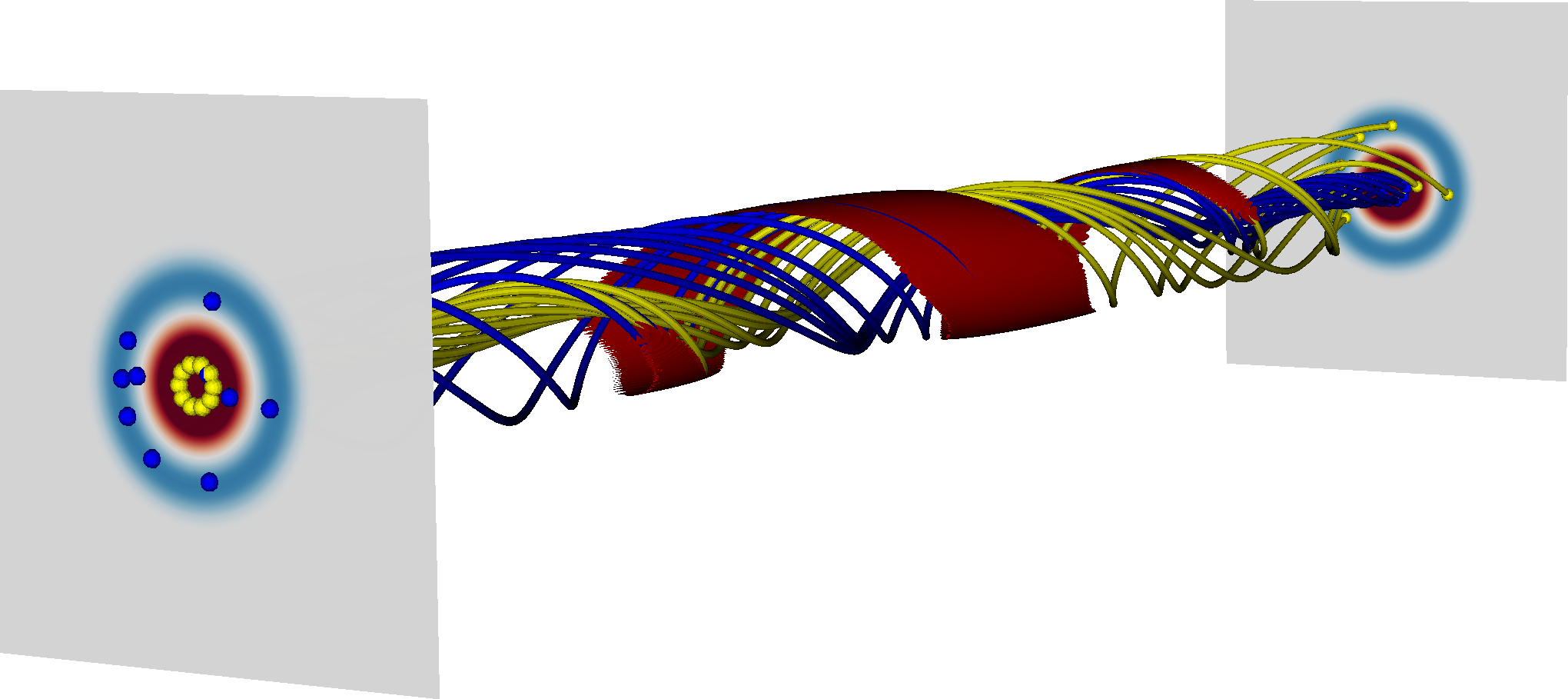}
      \put (50,40) {\small\textbf{(c)}}
    \end{overpic}
  \end{subfigure}
  \caption{\textbf{(a)}: \textit{Logarithmic plot of the total kinetic
      energy during the linear phase.} Overlaid is a \rs{straight}
    line corresponding to the linear growth rate $\sigma = 0.13$. The
    isotropic case is represented as a blue, solid line and the
    switching \rs{case} as an orange, dashed line. Though the kinetic energy is initially slightly greater using the switching \rs{model}, the growth rate appears unaffected by choice of viscosity model. The duration of the linear phase also appears to be negligibly affected.\\ \textbf{(b, c)}: \textit{The transition from linear to nonlinear instability in the isotropic case.} Times are \textbf{(b)} $t=45$ and \textbf{(c)} $t=50$. The yellow field lines start at $z=10$ and the blue field lines at $z=-10$. The isosurfaces are at $|\vec{j}| = 4$. The slices are plots of $\alpha(r)$. The linear growth of the instability ends around $t=35$ and the inner field compresses into the outer field, creating a current sheet. Between $t=45$ and $50$ this current sheet enables reconnection between the two regions. The transition for the switching case is qualitatively similar. In all three plots, the viscosity and resistivity are $\nu = 10^{-4}$ and $\eta = 5\times 10^{-4.5}$, respectively.}
  \label{fig:log_kinetic_energy_over_time}
\end{figure}

The linear development of the kink instability lasts until $t\approx
35$ \rs{as illustrated in
  Figure~\ref{fig:log_kinetic_energy_over_time}(a) and has} a
measured linear growth rate of $\sigma = 0.13$. Since the initial
velocity perturbation is calculated from an ideal and inviscid MHD
model with a piecewise constant $\alpha(r)$ in the equilibrium
configuration, the perturbation does not necessarily represent the
most unstable mode for the setup of the simulation. \rs{For this
reason} there is a brief transient period before the exponential rise of the instability at $t\approx10$, as shown in Figure~\ref{fig:log_kinetic_energy_over_time}(a). The isotropic model damps this initial velocity perturbation more than the switching model, leading to a small difference in kinetic energy during the growth of the linear instability, although the growth rate appears to be identical across the two models. The duration of the linear phase is also unaffected by the choice of viscosity model.

Initially, the instability occurs in the inner region of twist,
$r<0.5$, where the magnetic field kinks helically. This section of the
magnetic field compresses into the outer region, creating a current
sheet along the length of the tube \rs{as shown in} Figure~\ref{fig:log_kinetic_energy_over_time}(b). As the field continues to be compressed, it provides a magnetic pressure force that stalls the linear growth. The greater kinetic energy in the switching case leads to greater compression and thus a larger (though not notably stronger) current sheet. After this point, the growth of the kink instability is no longer in the linear phase.

During the transition from the linear to the nonlinear phase, field lines in the current sheet between the regions of inner and outer twist start to reconnect (Figure~\ref{fig:log_kinetic_energy_over_time}(b) and (c)). This happens sooner in the switching case, due to the larger compression.

\subsection{Nonlinear phase}

\begin{figure}[t]
    \centering
    \begin{subfigure}[t]{0.49\textwidth}
      \centering
      \begin{overpic}[]{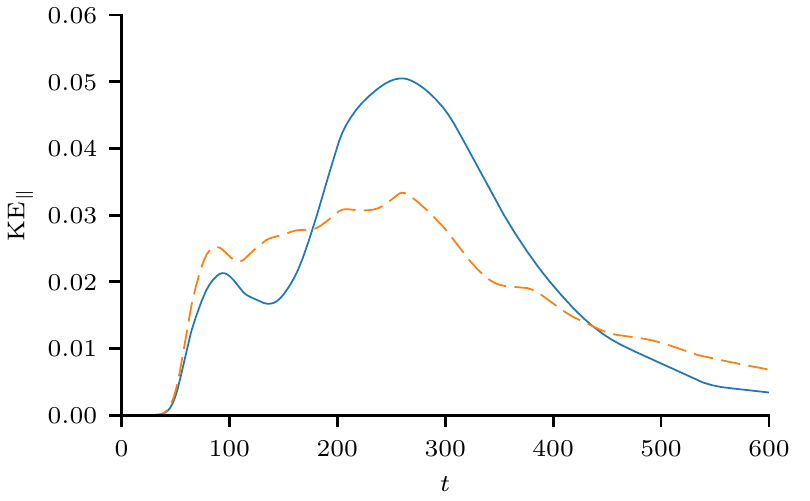}
        \put (50,55) {\small\textbf{(a)}}
      \end{overpic}
    \end{subfigure}%
    \begin{subfigure}[t]{0.49\textwidth}
      \centering
      \begin{overpic}[]{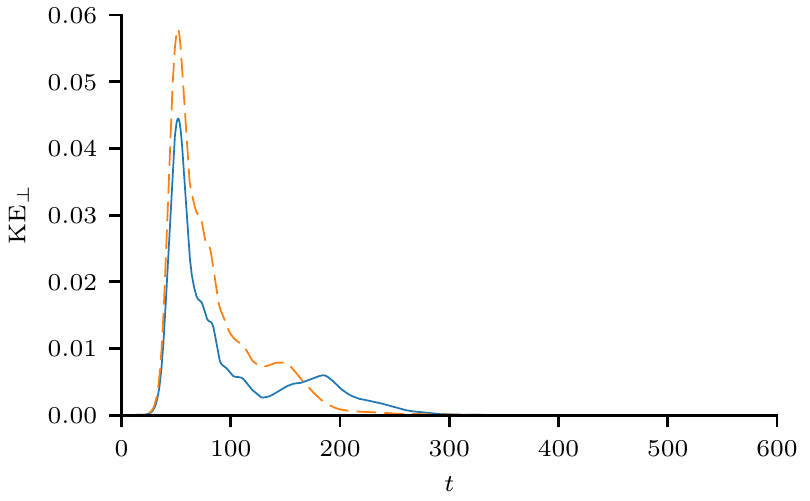}
        \put (50,55) {\small\textbf{(b)}}
      \end{overpic}
    \end{subfigure}
    \begin{subfigure}[t]{0.49\textwidth}
      \centering
      \begin{overpic}[]{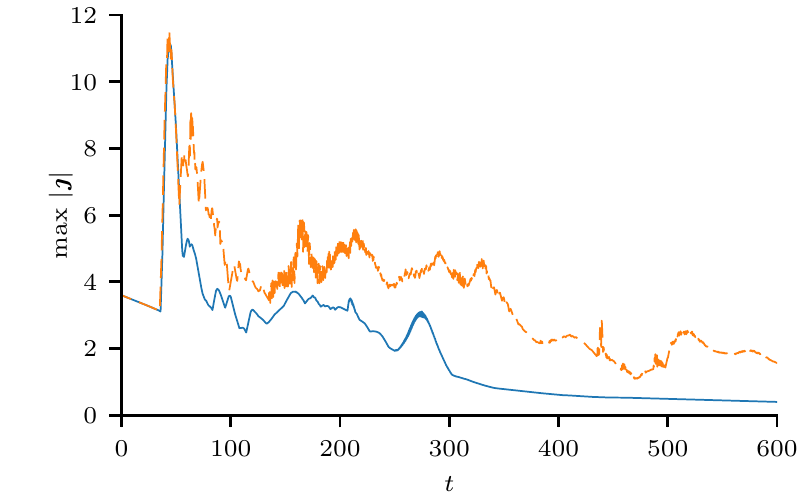}
        \put (50,55) {\small\textbf{(c)}}
      \end{overpic}
    \end{subfigure}
    \begin{subfigure}[t]{0.49\textwidth}
      \centering
      \begin{overpic}[width=\textwidth]{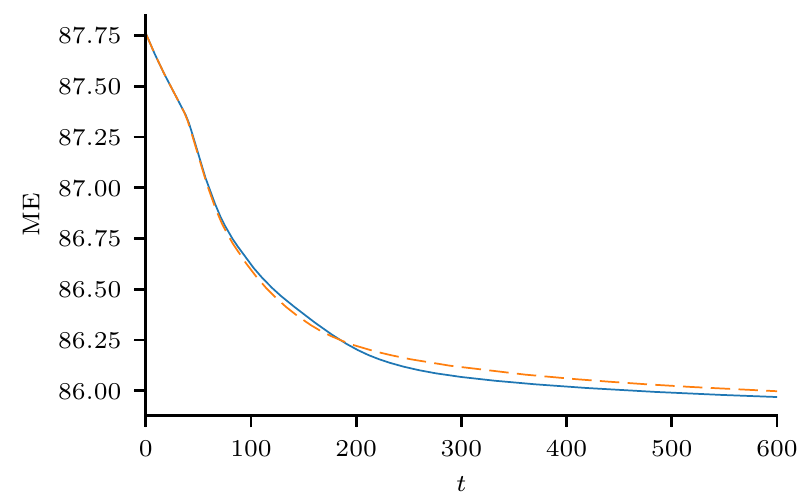}
        \put (50,55) {\small\textbf{(d)}}
      \end{overpic}
    \end{subfigure}
    \caption{\textit{\rs{Energy components and current} as functions of time.} \textbf{(a)} Parallel kinetic energy, \textbf{(b)} perpendicular kinetic energy, \textbf{(c)} maximum current density and \textbf{(d)} magnetic energy density as functions of time for isotropic (blue, solid) and switching (orange, dashed) viscosity, with diffusion parameters $\nu = 10^{-4}$ and $\eta = 5\times 10^{-4.5}$.}
    \label{fig:energies}
\end{figure}

Although the choice of viscosity model has a small effect on the
linear phase of the kink instability, it does play an important role
in the development of the nonlinear phase. By examining the kinetic
energies (KEs) in Figures~\ref{fig:energies}(a) and (b), a pattern
emerges in both cases that \rr{has similarities with the nonlinear
  behaviour of kink instabilities described in} Hood et
al.~\cite{hoodCoronalHeatingMagnetic2009}. Shortly after the linear
phase, at $t\approx50$, the KEs for both viscosity models
\rs{exhibits} a sharp rise, with the KEs associated with the switching
model attaining higher amplitudes. At the same time, a sharp rise is
also found in the maximum current \rs{as seen in}
Figure~\ref{fig:energies}(c) and, leading on from this spike, the
current magnitudes associated with the switching model are larger than
those associated with the isotropic model. Returning to the KEs, the
energies associated with the switching model are greater until
$t\approx175$, after which, in \emph{only} the isotropic case, we see
a clear secondary spike in perpendicular kinetic energy, and a large
increase in parallel kinetic energy, much greater than the energy
shown in the switching case. It is difficult to detect this new phase
in the maximum current (Figure~\ref{fig:energies}(c)), but it is found
in other quantities related to magnetic reconnection. 

\begin{figure}[t]
    \centering
    \begin{subfigure}[t]{0.5\textwidth}
      \centering
      \begin{overpic}[width=\linewidth]{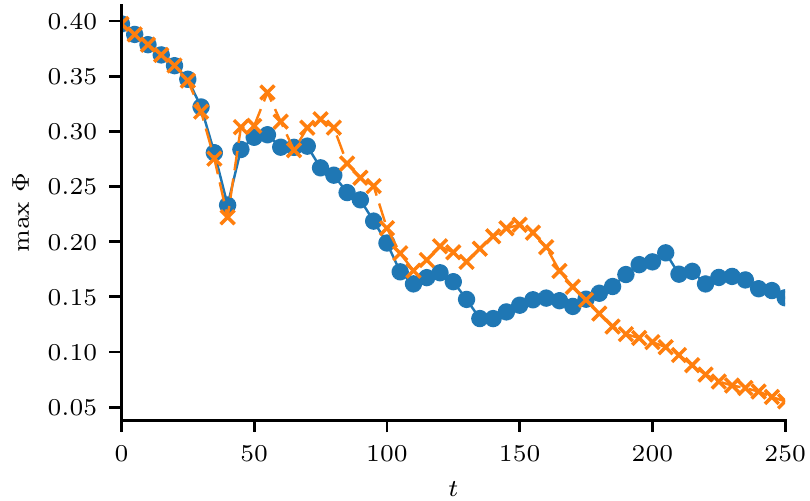}
        \put (50,60) {\small\textbf{(a)}}
      \end{overpic}
      \label{fig:max_parallel_electric_field}
    \end{subfigure}%
    ~
    \begin{subfigure}[t]{0.5\textwidth}
      \centering
      \begin{overpic}[width=\linewidth]{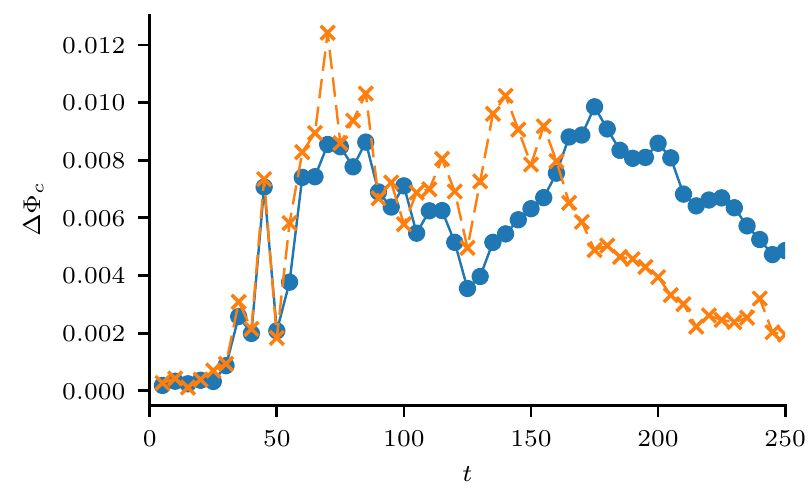}
        \put (50,60) {\small\textbf{(b)}}
      \end{overpic}
      \label{fig:mean_difference_in_connectivity}
    \end{subfigure}
    \caption{\textit{Reconnection rates.} \textbf{(a)} The maximum
      integrated parallel electric field and \textbf{(b)} the mean
      difference in connectivity using isotropic viscosity (blue dot
      \& solid line) and switching viscosity (orange cross \& dashed
      line) with $\nu = 10^{-4}$ and $\eta = 5\times 10^{-4.5}$. \rs{A
        cadence of} $5$ Alfv\'en times \rs{is used} between data points.}
    \label{fig:reconnection-rates}
\end{figure}

Figure~\ref{fig:reconnection-rates} displays the time series of the maximum integrated parallel electric field and the mean difference in connectivity, for both viscosity models. \rt{Both of the time series in Figure~\ref{fig:reconnection-rates} display similar trends to those found in the perpendicular kinetic energy plot, Figure~\ref{fig:energies}(a). For both isotropic and anisotropic viscosity there appears to be two major peaks in the reconnection measures that align with peaks in the perpendicular kinetic energy. This is much more obvious in the isotropic case. We can conclude that both viscosity models allow for two phases of reconnection but the time at which they occur is significantly modified by the form of viscosity chosen.} It is, therefore, clear that the form of viscosity is having a significant effect on the nonlinear evolution of the kink instability, both on the flow dynamics and the reconnection of the magnetic field. We will now examine, in more detail, the two important phases indicated by the isotropic time series, and how the switching case differs.

\subsection{First phase: $t\approx65$--$100$}

\begin{figure}[t]
  \centering
  \begin{subfigure}[b]{0.32\textwidth}
  \begin{center}
      \begin{overpic}[width=\textwidth]{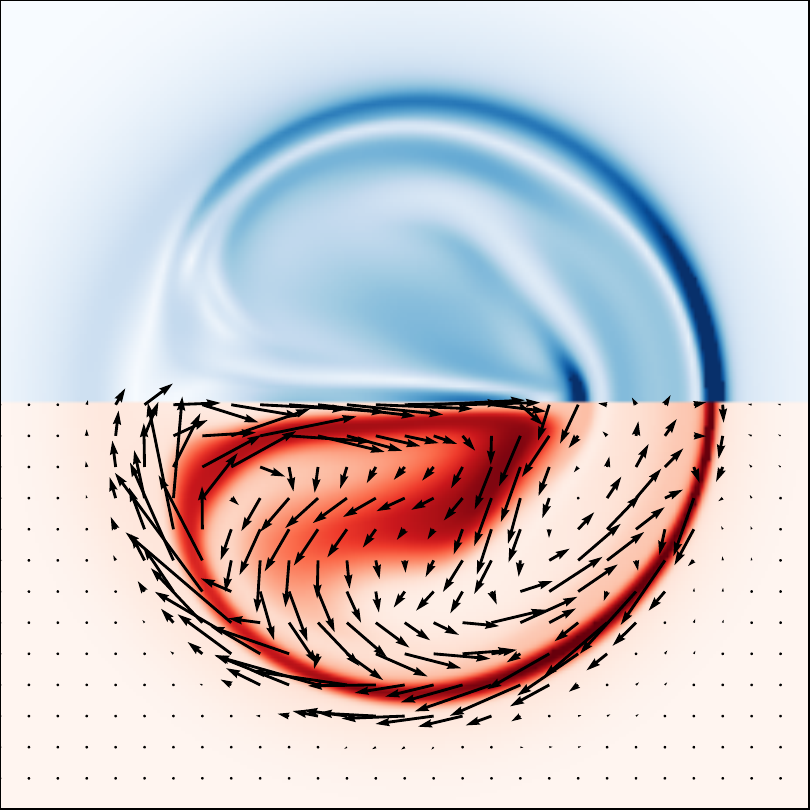}
        \put (3,92) {\small\textbf{(a)}}
      \end{overpic}
  \end{center}
  \end{subfigure}
  \begin{subfigure}[b]{0.32\textwidth}
  \begin{center}
      \begin{overpic}[width=\textwidth]{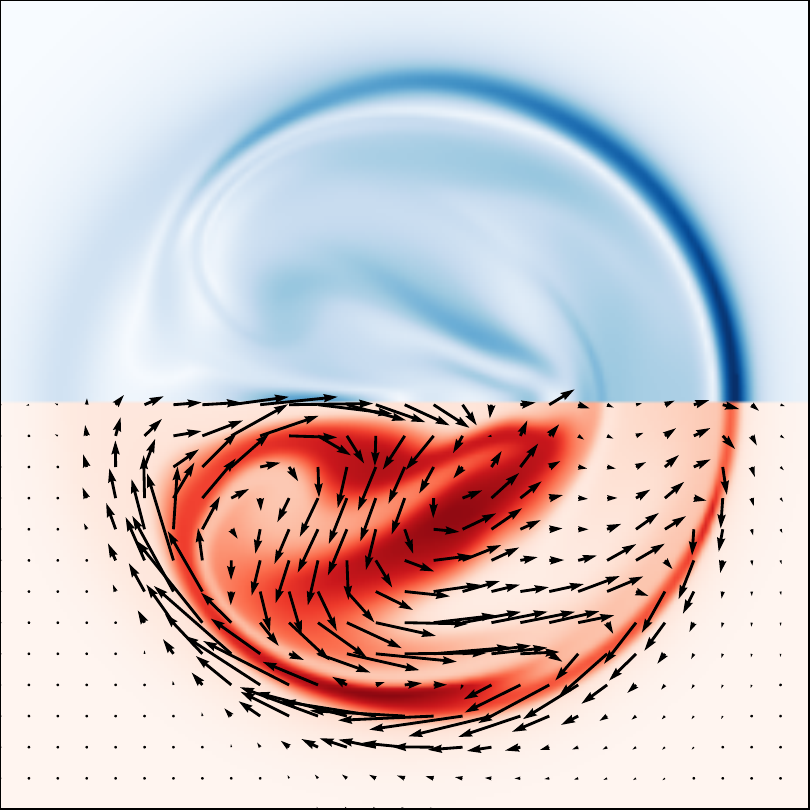}
        \put (3,92) {\small\textbf{(b)}}
      \end{overpic}
  \end{center}
  \end{subfigure}
  \begin{subfigure}[b]{0.32\textwidth}
  \begin{center}
      \begin{overpic}[width=\textwidth]{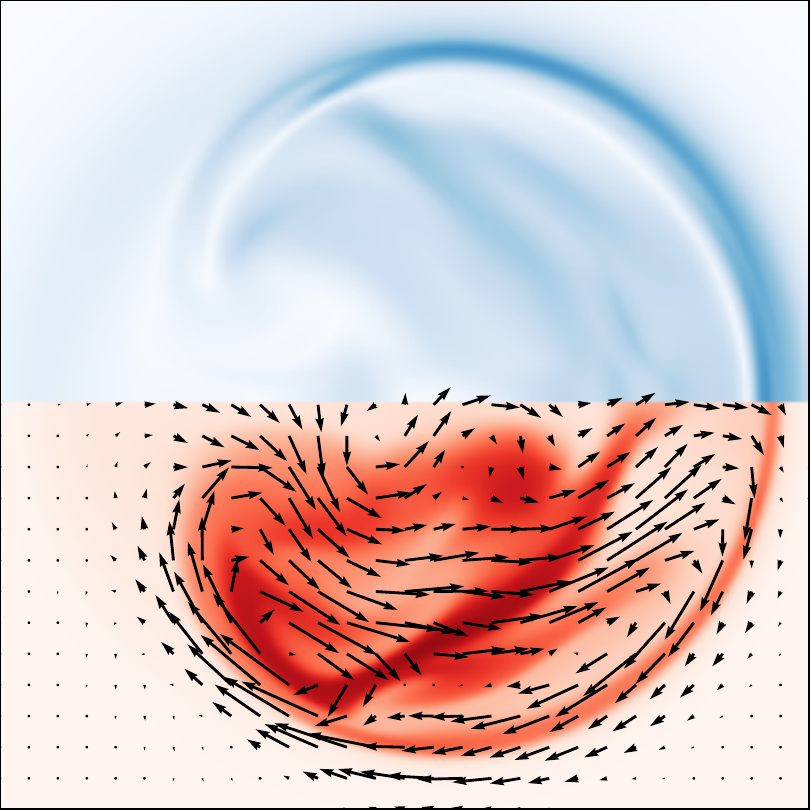}
        \put (3,92) {\small\textbf{(c)}}
      \end{overpic}
  \end{center}
  \end{subfigure}
  \begin{subfigure}[b]{0.32\textwidth}
  \begin{center}
      \begin{overpic}[width=\textwidth]{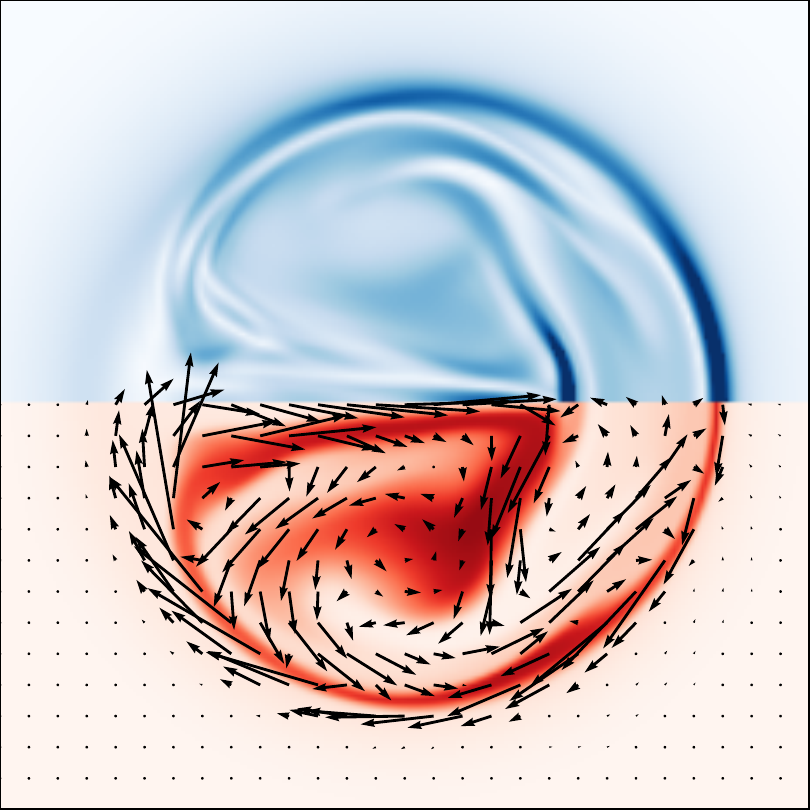}
        \put (3,92) {\small\textbf{(d)}}
      \end{overpic}
  \end{center}
  \end{subfigure}
  \begin{subfigure}[b]{0.32\textwidth}
  \begin{center}
      \begin{overpic}[width=\textwidth]{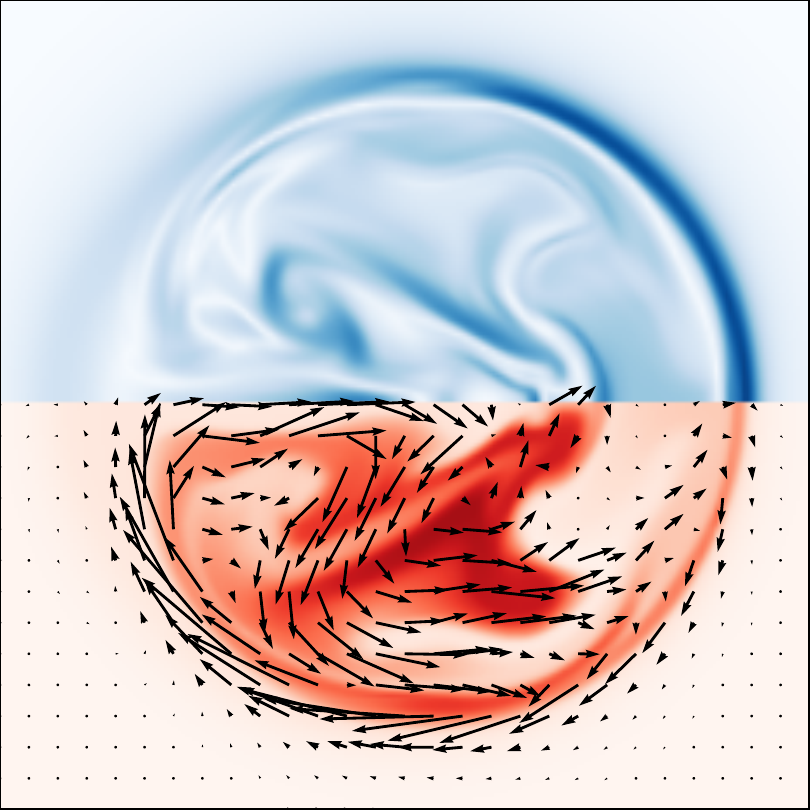}
        \put (3,92) {\small\textbf{(e)}}
      \end{overpic}
  \end{center}
  \end{subfigure}
  \begin{subfigure}[b]{0.32\textwidth}
  \begin{center}
      \begin{overpic}[width=\textwidth]{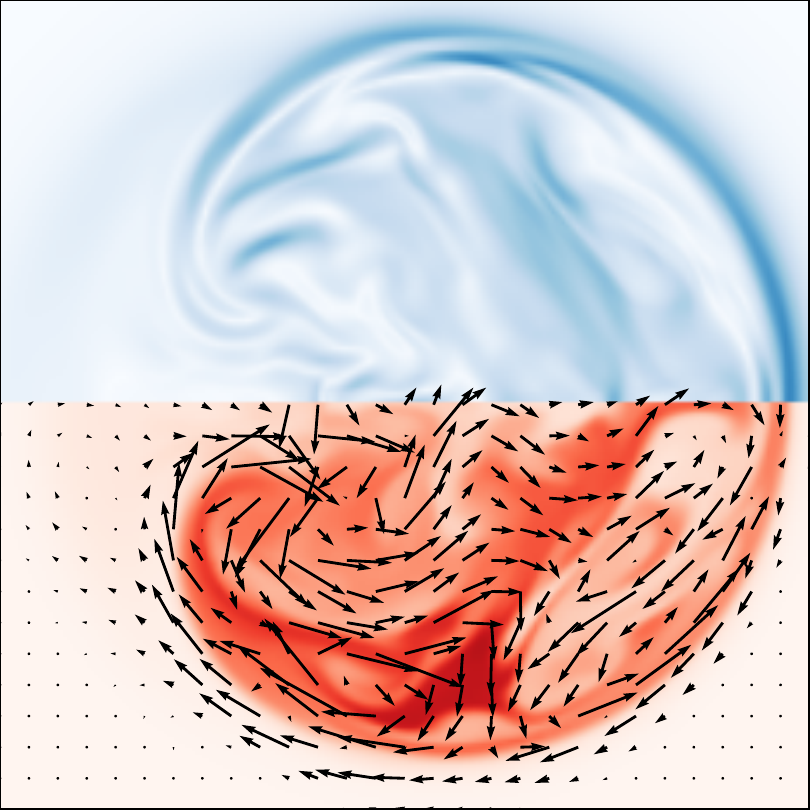}
        \put (3,92) {\small\textbf{(f)}}
      \end{overpic}
  \end{center}
  \end{subfigure}
  \caption{\textit{The difference in the evolution of current density,
      temperature and velocity structures \rs{between the isotropic
        and the switching viscosity cases}.} Slices at $z=0$ of
    current density (top of each figure; blue is $|\vec{\jmath}| =
    3.5$, white is $|\vec{\jmath}| = 0$) and temperature (bottom of
    each figure; red is $T = 5\times10^{-2}$, white is
    $T=1.15\times10^{-5}$), overlaid with fluid flow, at times $t=65$
    (left), $75$ (middle), and $100$ (right). \ro{The halves shown are identical to their counterparts that are not shown, for both temperature and current density. That is, the simulation is vertically symmetrical at these times.} The profile is cropped to
    $x=\pm1,\ y=\pm1$. The top three \rs{panels (a) to (c)} show the
    isotropic case, the bottom three \rs{panels (d) to (f) show} the switching case.}
  \label{fig:turning-point}
\end{figure}

At $t=65$, an intense current structure appears near the centre of the
tube for both viscosity models, although it is much stronger in the
switching case \rs{as illustrated in Figure~\ref{fig:turning-point}}. Since the viscous damping associated with parallel viscosity is much less than that of isotropic viscosity, the flows in the switching case are stronger than those in the isotropic case (Figures~\ref{fig:energies}(a) and (b)). The faster flows drive stronger reconnection in the central current structure (see Figure~\ref{fig:reconnection-rates}) and the interaction of these processes leads to stronger outflows and finer-scale structures in the switching model case compared with the isotropic model case. Evidence of this behaviour can be seen by comparing the current and flow structures in Figure~\ref{fig:turning-point}. The effects of this phase can also be seen in the magnetic energy evolution, shown in Figure~\ref{fig:energies}(d). Between times $t=100$ and $125$, due to stronger reconnection in the switching case, the magnetic field relaxes \rt{marginally faster than that of the isotropic case, before the secondary instability begins in the isotropic case around $t=125$.}

\begin{figure}[t]
  \centering
  \begin{subfigure}[b]{0.32\textwidth}
  \begin{center}
      \begin{overpic}[width=\textwidth]{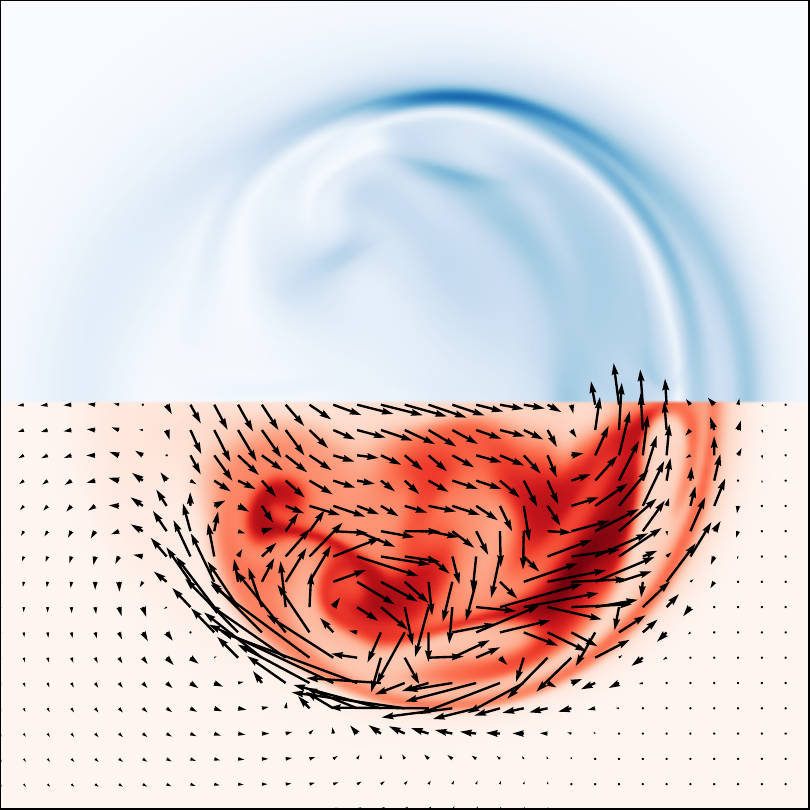}
        \put (3,92) {\small\textbf{(a)}}
      \end{overpic}
  \end{center}
  \end{subfigure}
  \begin{subfigure}[b]{0.32\textwidth}
  \begin{center}
      \begin{overpic}[width=\textwidth]{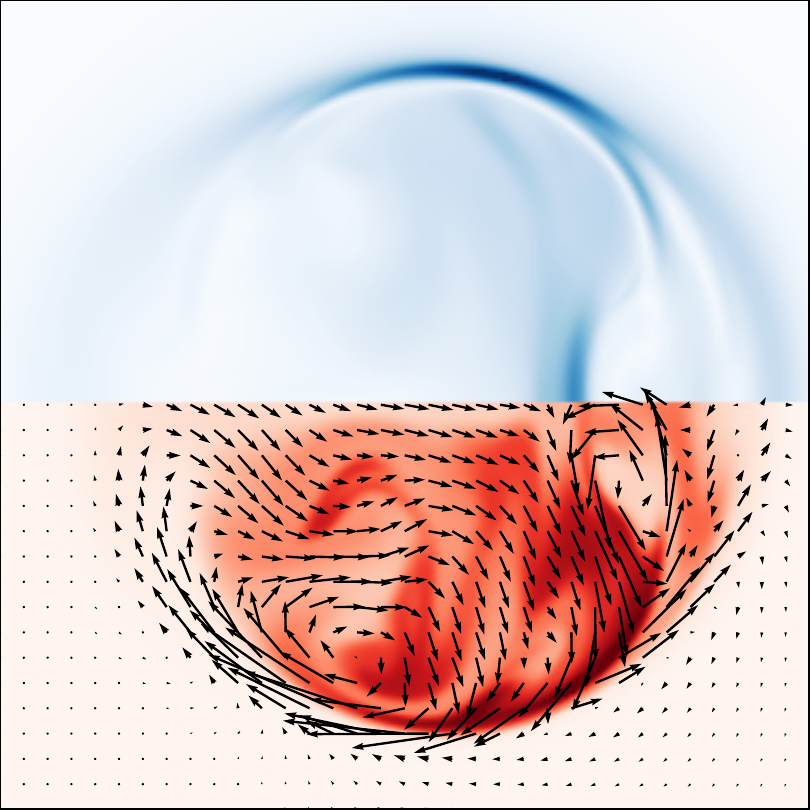}
        \put (3,92) {\small\textbf{(b)}}
      \end{overpic}
  \end{center}
  \end{subfigure}
  \begin{subfigure}[b]{0.32\textwidth}
  \begin{center}
      \begin{overpic}[width=\textwidth]{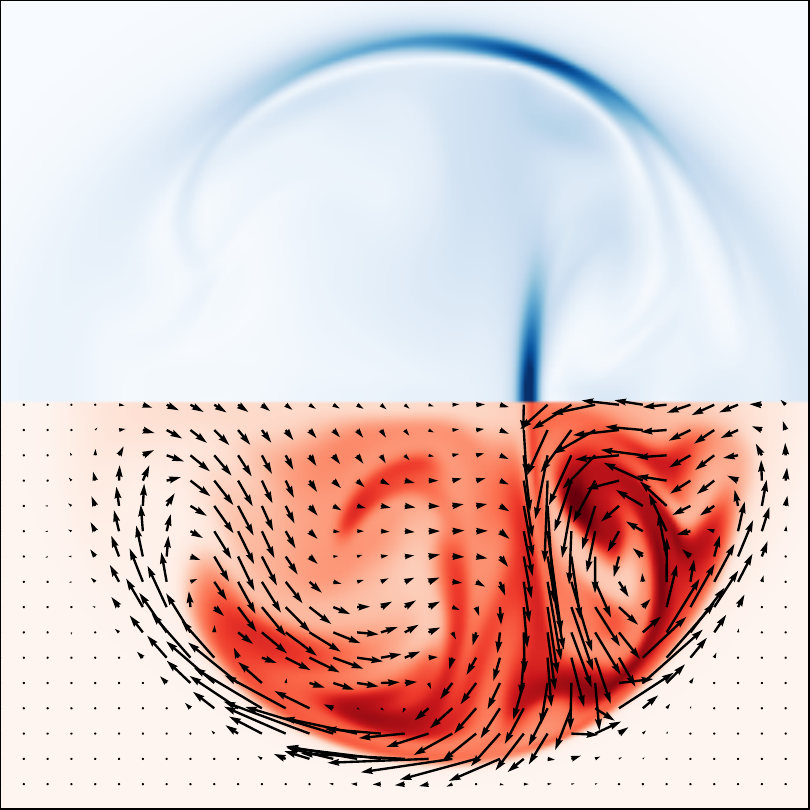}
        \put (3,92) {\small\textbf{(c)}}
      \end{overpic}
  \end{center}
  \end{subfigure}
  \begin{subfigure}[b]{0.32\textwidth}
  \begin{center}
      \begin{overpic}[width=\textwidth]{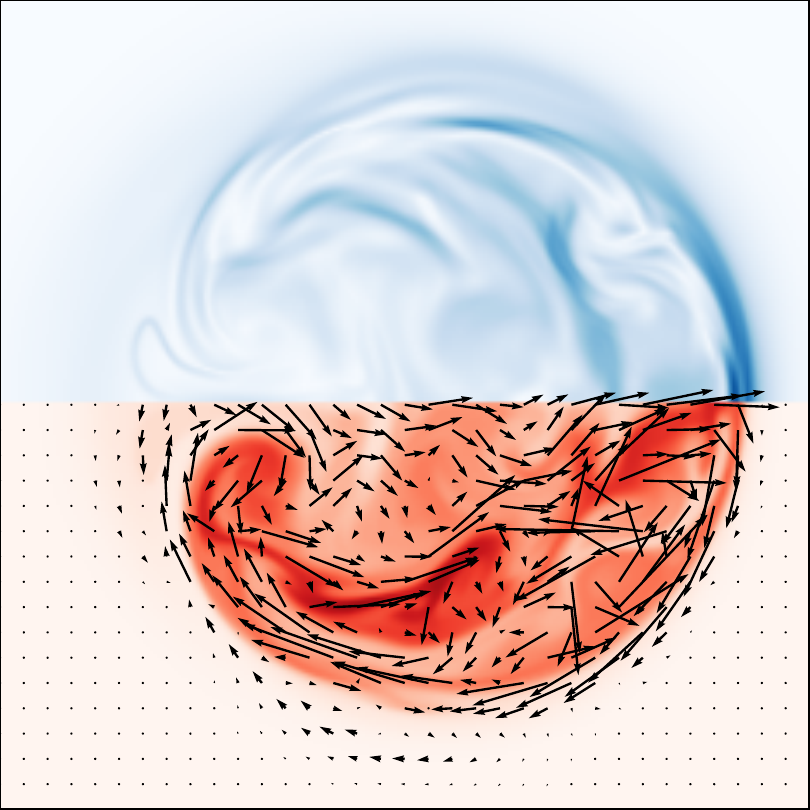}
        \put (3,92) {\small\textbf{(d)}}
      \end{overpic}
  \end{center}
  \end{subfigure}
  \begin{subfigure}[b]{0.32\textwidth}
  \begin{center}
      \begin{overpic}[width=\textwidth]{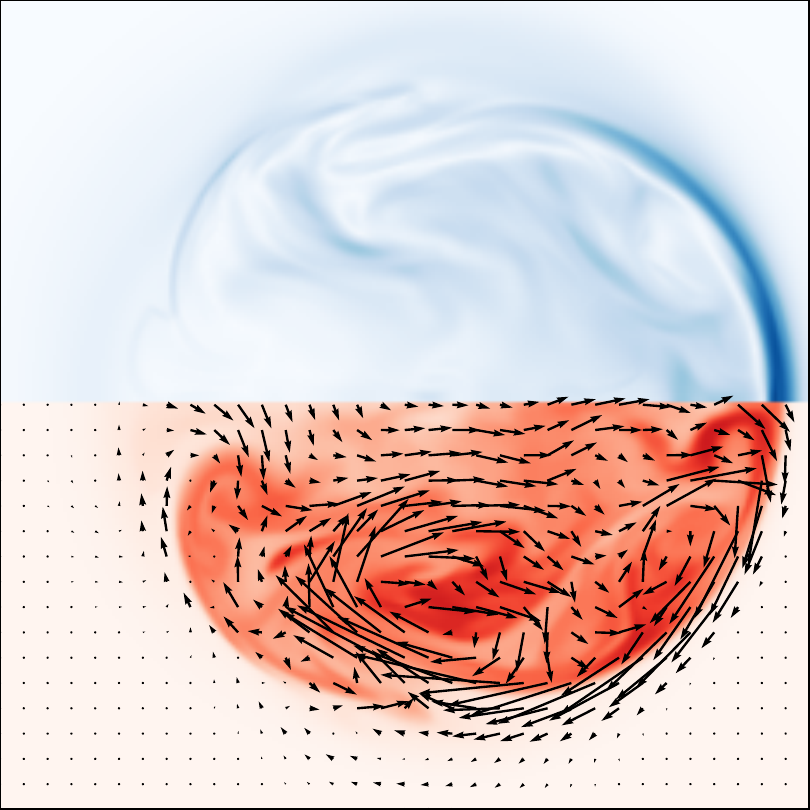}
        \put (3,92) {\small\textbf{(e)}}
      \end{overpic}
  \end{center}
  \end{subfigure}
  \begin{subfigure}[b]{0.32\textwidth}
  \begin{center}
      \begin{overpic}[width=\textwidth]{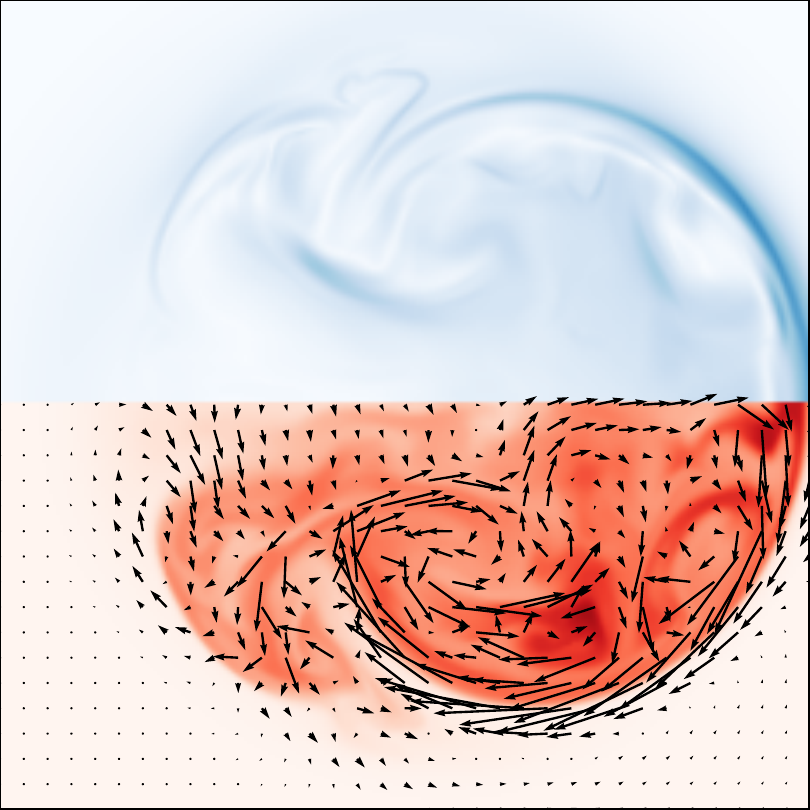}
        \put (3,92) {\small\textbf{(f)}}
      \end{overpic}
  \end{center}
  \end{subfigure}
  \caption{\textit{The formation of a reconnection feedback loop
\rs{in the isotropic and the switching viscosity cases}.} Slices at
    $z=0$ of current density (top of each figure; blue is
    $|\vec{\jmath}| = 3$, white is $|\vec{\jmath}| = 0$) and
    temperature (bottom of each figure; red is $T = 4.9 \times
    10^{-2}$, white is $T=1.15\times 10^{-5}$), overlaid with fluid
    flow, at times $t=125$ (left), $150$ (middle), and $175$
    (right). The structure is nearly exactly symmetrical. The profile
    is cropped to $x=\pm1.2,\ y=\pm1.2$. The top three \rs{panels (a)
      to (c)} show the isotropic case, the bottom three \rs{panels (d)
      to (f) show} the switching case. The isotropic case shows two current sheets causing reconnection at the top and bottom of the tube, producing flows that sustains another central current sheet, which feeds back into the top and bottom sheets. The switching case instead shows one single main current sheet at the right hand side, along with numerous smaller current structures throughout the domain.}
  \label{fig:feedback-reconnection}
\end{figure}

\subsection{Second phase: $t\approx125$--$175$}
The contrast between fine-scale current and flow structures for the switching model, and the smoother, larger-scale structures of the isotropic model continues to be present at later times. Figure~\ref{fig:feedback-reconnection} shows the same data as Figure~\ref{fig:turning-point} but for the times $t=125$, $150$ and $175$. Looking at the slices for $t=125$, there is more fine-scale structure generated in the switching case compared to the isotropic case, as in the first phase described above. This second phase, however, marks the beginning of a significant change in behaviour in the isotropic model case. From Figure~\ref{fig:energies}, the parallel KE for the isotropic model exhibits a rapid and large increase in kinetic energy, characteristic of a secondary instability. \rt{To a lesser extent, there is also growth in the perpendicular KE, and the two reconnection measures for the isotropic model. In the second phase, these three measures increase to eventually become greater than their corresponding values for the switching model, around $t=175$.} In order to understand this significant difference in the behaviour between the two models, we will first consider the slices in Figure~\ref{fig:feedback-reconnection}.

\begin{figure}[t]
  \centering
  \begin{subfigure}[b]{0.48\textwidth}
  \begin{center}
    \begin{overpic}[width=\textwidth]{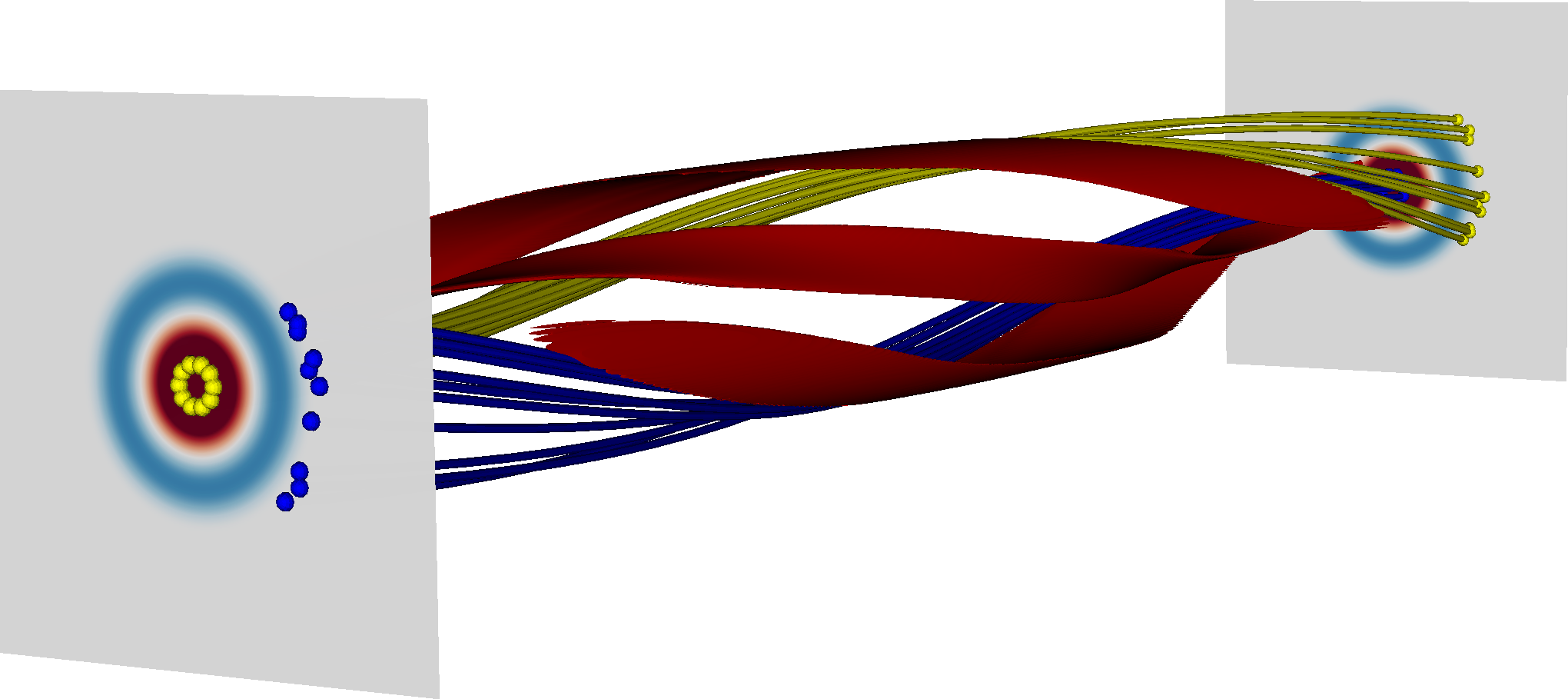}
      \put (50,40) {\small\textbf{(a)}}
    \end{overpic}
  \end{center}
  \end{subfigure}
  \begin{subfigure}[b]{0.48\textwidth}
  \begin{center}
    \begin{overpic}[width=\textwidth]{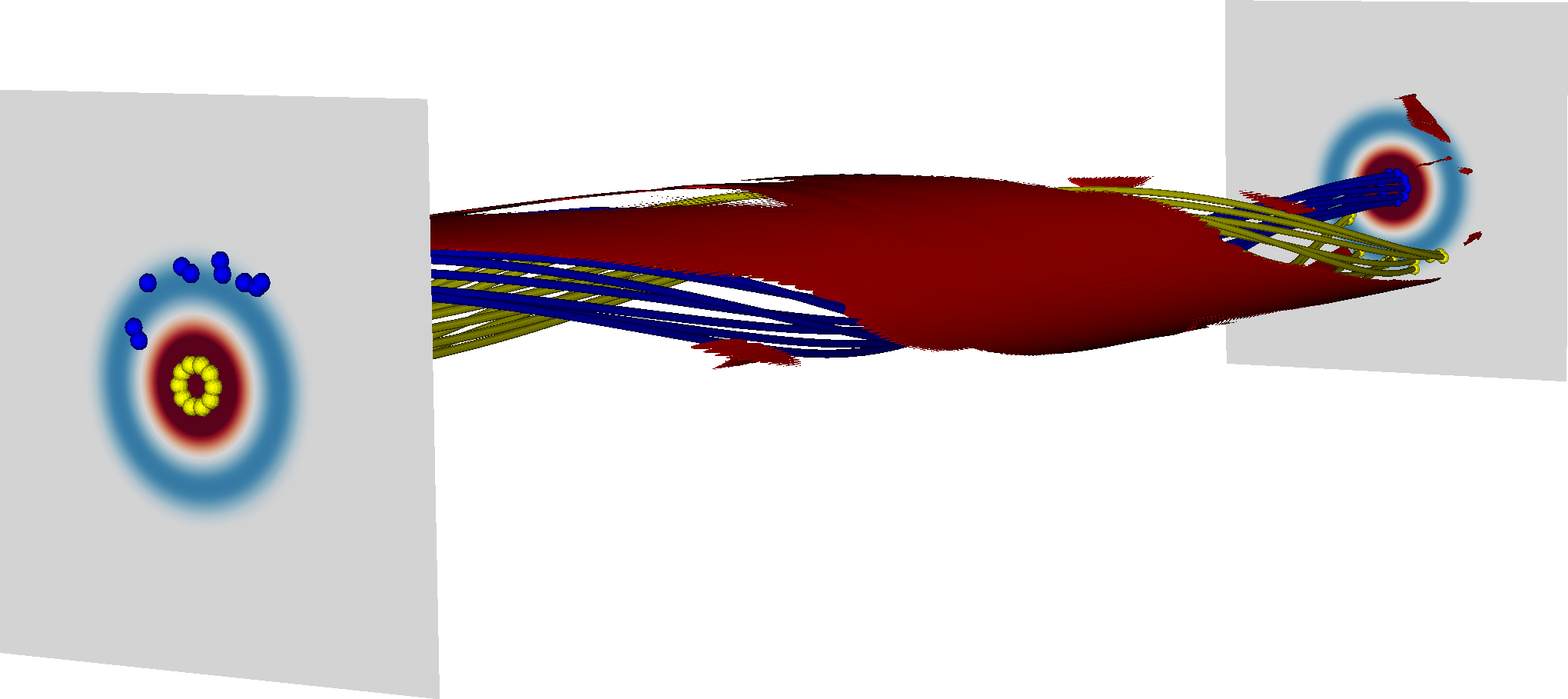}
      \put (50,40) {\small\textbf{(b)}}
    \end{overpic}
  \end{center}
  \end{subfigure}
  \caption{\textit{The difference in 3D current structures.} The two different current structures in \textbf{(a)} the isotropic case and \textbf{(b)} the switching case at $t=175$. Isosurfaces are at $|\vec{\jmath}| = 1.5$.}
\label{fig:reconnection-field-lines}
\end{figure}

At $t=125$ (panels (a) and (d) \rs{of Figure~\ref{fig:feedback-reconnection}}), the difference in behaviour
between the models is similar to the first phase but some new features
appear. The KE in the isotropic model begins to increase and, as
mentioned before, appears to signify a secondary instability. In
Figure~\ref{fig:feedback-reconnection}(a), two new current sheets have
formed at the top and bottom of the tube. A three-dimensional (3D)
visualisation of these current sheets is shown in
Figure~\ref{fig:reconnection-field-lines}(a). The outflow from the
reconnection occurring within these current sheets then creates two
new symmetric vortices on the right hand side of the tube, advecting
the field into the centre of the tube. This behaviour can be seen
clearly in \rs{Figure~\ref{fig:feedback-reconnection}}(b) where vortex
motion compresses the magnetic field and forms a central region of
enhanced current density. Later, as seen at $t=175$ in \rs{Figure~\ref{fig:feedback-reconnection}(c)}, the central current region becomes stronger due to continued compression and reconnection ensues, becoming stronger than the switching model case (see Figure~\ref{fig:reconnection-rates}). The outflows from this current region then feed into the vortical motions that drive the compression. In this way, a feedback loop is set up, and the reconnection within the current structure continuously drives the flow, resulting in an instability. \ro{This kind of interaction between multiple current sheets is also seen in~\cite{hoodCoronalHeatingMagnetic2009}.} Due to this secondary instability, magnetic relaxation now becomes faster for the isotropic case. The magnetic energy for this case now dips below that of the switching case, as shown in Figure~\ref{fig:energies}(d).

During this phase, the kinetic energy in the switching model case also increases but to a much smaller extent compared to the isotropic model case. Although the current densities in Figures~\ref{fig:feedback-reconnection}\rs{(d) to (f)} again exhibit finer-scale structure compared to the isotropic case, the magnitude of the current density within the tube becomes weaker with a more uniform profile developing in time. The dominating current sheets are on the edge of the tube, as also indicated in Figure~\ref{fig:reconnection-field-lines}(b).

\subsection{Late-time states}

\begin{figure}[t]
  \centering
  \begin{subfigure}[b]{0.48\textwidth}
  \begin{center}
    \begin{overpic}[width=\textwidth]{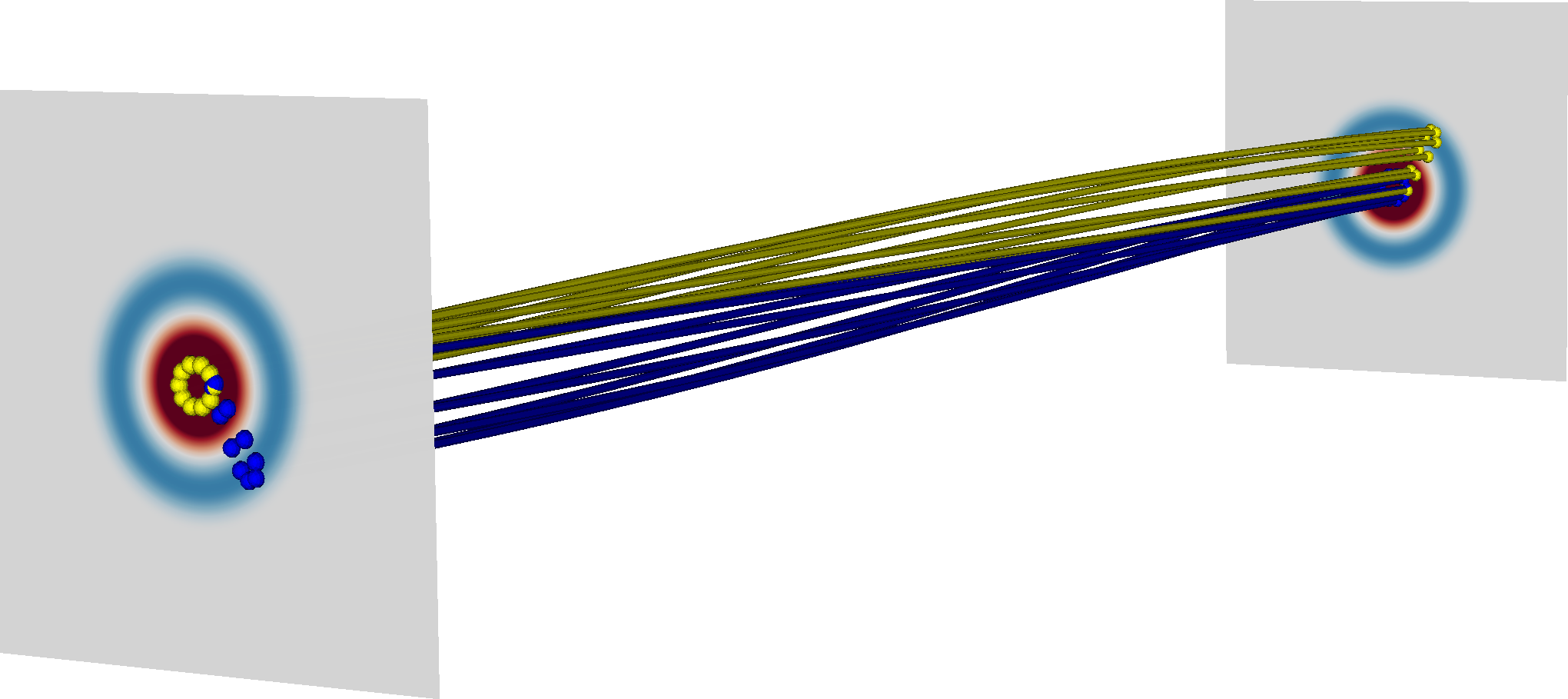}
      \put (50,40) {\small\textbf{(a)}}
    \end{overpic}
  \end{center}
  \end{subfigure}
  \begin{subfigure}[b]{0.48\textwidth}
  \begin{center}
    \begin{overpic}[width=\textwidth]{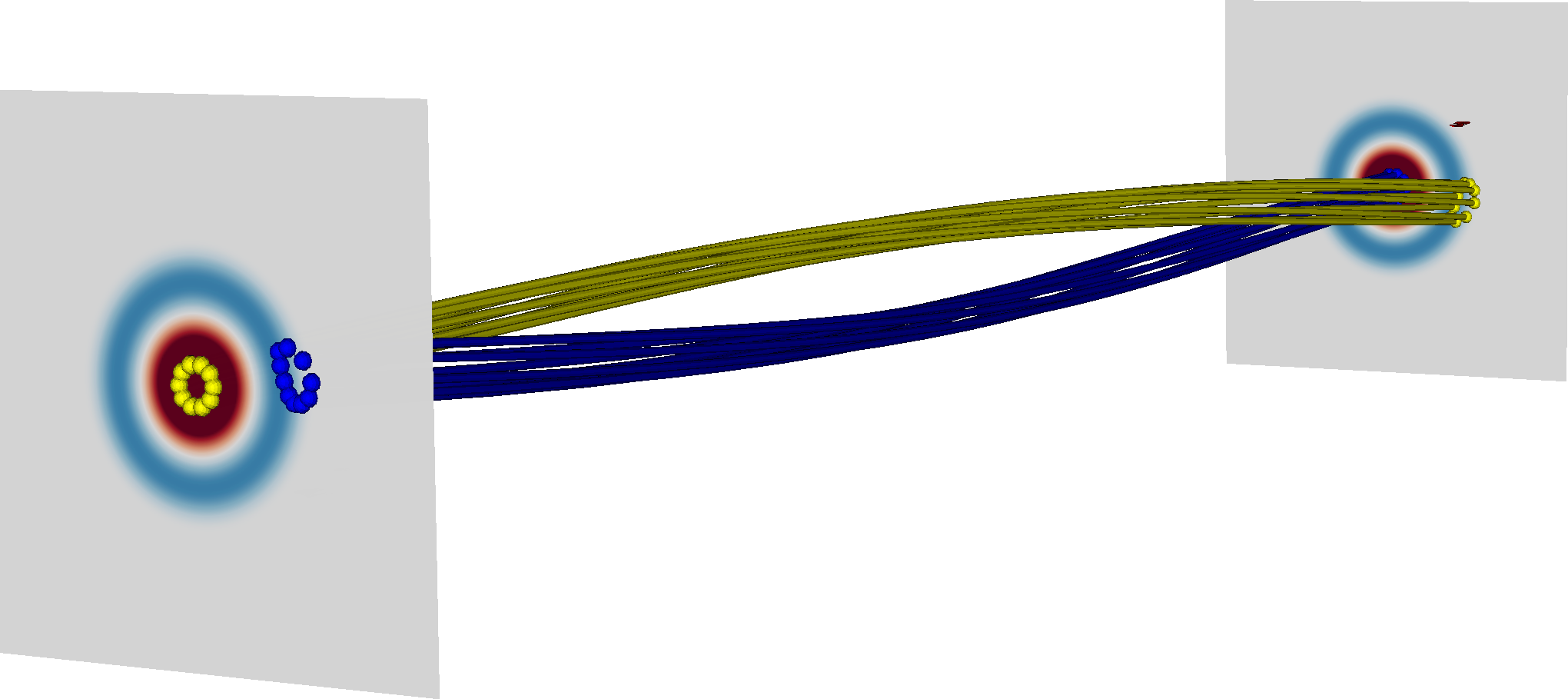}
      \put (50,40) {\small\textbf{(b)}}
    \end{overpic}
  \end{center}
  \end{subfigure}
  \caption{\textit{Late-time magnetic field structures.} Magnetic field lines plotted at $z=\pm10$ in \textbf{(a)} the isotropic case and \textbf{(b)} the switching case at $t=600$.}
\label{fig:finale-field-lines}
\end{figure}

For both cases, the asymptotic relaxed magnetic field is a \rt{linear force-free field}. The route to this asymptotic state, however, depends on the viscosity model used. At the late time of $t=600$, there remain clear differences in the field structure between the two models resulting from the different nonlinear evolutions, as can be seen in Figure~\ref{fig:finale-field-lines}. \rr{At $t=600$, the magnetic} field in the isotropic case (Figure~\ref{fig:finale-field-lines}(a)) appears straighter, indicative of \rr{more efficient magnetic relaxation}. Indeed, Figure~\ref{fig:energies}(d) shows that more energy has been extracted from the field in the isotropic case. At $t=600$, the current density and energies (see Figure~\ref{fig:energies}) are still non-zero, so further relaxation is expected. For coronal applications, however, these late times are not as important as the early phases, described above, when the initial and secondary instabilities develop.

\subsection{Viscous and Ohmic heating}

\begin{figure}[t]
    \centering
    \begin{subfigure}[t]{0.32\textwidth}
      \centering
      \begin{overpic}[width=\textwidth]{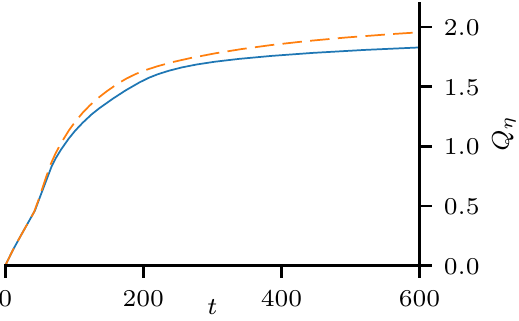}
        \put (40,60) {\small\textbf{(a)}}
      \end{overpic}
    \end{subfigure}
    \begin{subfigure}[t]{0.32\textwidth}
      \centering
      \begin{overpic}[width=\textwidth]{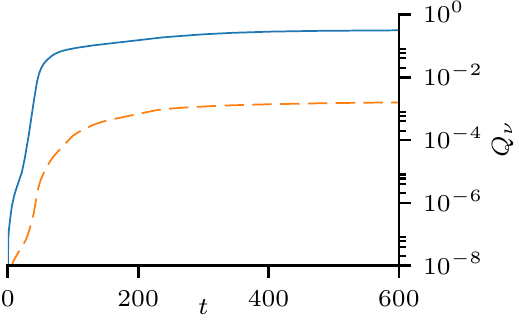}
        \put (40,60) {\small\textbf{(b)}}
      \end{overpic}
    \end{subfigure}
    \begin{subfigure}[t]{0.32\textwidth}
      \centering
      \begin{overpic}[width=\textwidth]{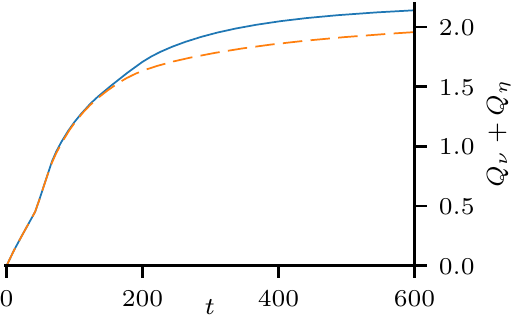}
        \put (40,60) {\small\textbf{(c)}}
      \end{overpic}
    \end{subfigure}
    \caption{\textit{Heating rates as functions of time.} \textbf{(a)}
      Ohmic, \textbf{(b)} viscous and \textbf{(c)} combined heating,
      for isotropic (blue, solid) and switching (orange, dashed)
      viscosity, with diffusion parameters $\nu = 10^{-4}$ and $\eta =
      5\times 10^{-4.5}$. Ohmic heating dominates isotropic viscous
      heating by an order of magnitude, and switching viscosity by
      four orders. Isotropic viscosity generates a factor of around
      $10^{3}$ more heat that switching viscosity. \rs{Even though} more Ohmic heat is generated in the switching case, it does not \rs{compensate} for the much weaker viscous heating.}
    \label{fig:heating}
\end{figure}

Over the lifetime of the entire instability the switching model allows for the generation of more Ohmic heating (Figure~\ref{fig:heating}(a)). This is despite the long, secondary phase of reconnection produced in the isotropic case. The greater heating in the switching case is due to two factors: the greater compression created by faster flows, creating stronger or larger current sheets and the more numerous current sheets created by more complex flows. However, isotropic viscous heating dominates that of the switching model by two orders of magnitude (Figure~\ref{fig:heating}(b)) ultimately leading to greater overall heating in the isotropic case (Figure~\ref{fig:heating}(c)). \rt{Physically, this is due to anisotropic viscosity only performing significant damping when velocity gradients align appropriately with the magnetic field (that is, when $(\ten{W} \vec{b}) \cdot \vec{b}$ is non-zero).} 

Comparing Ohmic and viscous heating (Figures~\ref{fig:heating}(a) and (b)), Ohmic heating outperforms viscous heating in both cases, by an order of magnitude in the isotropic case and by three orders in the switching case. \rt{Even though we use similar values for the diffusion of the magnetic field $\eta$ and the velocity $\nu$, during the kink instability the current sheets produced are much stronger than the gradients in velocity, hence the Ohmic heating dissipates more energy than the viscous heating.}

\ro{Due to the relationship between $(\ten{W} \vec{b}) \cdot \vec{b}$ and $Q_{\nu}$ (equation~\eqref{eq:aniso_viscous_heating} with $s\approx 1$), the small magnitude of $Q_{\nu}$ in Figure~\ref{fig:heating}(b) implies that $(\ten{W} \vec{b}) \cdot \vec{b}$ is small everywhere. With the anisotropic viscous heating being heavily dependent on the magnetic field direction and since $(\ten{W} \vec{b}) \cdot \vec{b}$ is small everywhere in the kink simulation, it follows that the anisotropic viscous heating is always lower in magnitude compared to the isotropic viscous heating, which is not bound by the diection of the magnetic field.}

\subsection{The effect of anisotropy on feedback reconnection}

We have described the nonlinear evolution of the kink instability for the cases of only isotropic viscosity and (practically) only anisotropic viscosity. In order to determine how much of each extreme form of viscosity controls the evolution of the secondary instability, \ro{we can fix the interpolation between isotropic and anisotropic viscosity by prescribing the switching function $s$ as a constant in equation~\eqref{eq:switching_stress_tensor}, instead of letting $s$ rely on the local field strength $|\vec{B}|$.} \rt{It should be noted that the simulations in which we fix $s$ are no longer physically realistic, but we use them only to estimate the amount of anisotropy required in the viscosity to disrupt the secondary instability.} Since the interpolation involves only $s^2$ instead of $s$, in practice we fix the value of $s^2$.

\begin{figure}[t]
  \centering
  \includegraphics[width=0.5\linewidth]{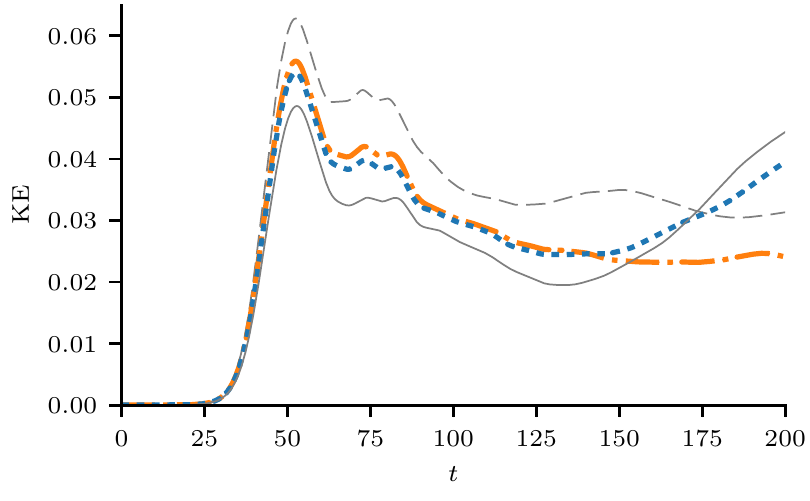}
  \caption{\textit{Kinetic energy over time, varying \rs{the switching
        function} $s^2$.} The grey lines are the two regular cases; switching, where $s^2 = 1$ (dashed), and isotropic (solid), where $s^2=0$. The coloured lines represent values of $s^2 = 0.5$ (blue, dotted), and $s^2 = 0.6$ (orange, dash-dotted). There is a clear critical value somewhere between $0.5$ and $0.6$, where the behaviour changes.}
  \label{fig:kinetic-energy-changing-s}
\end{figure}

By letting $s^2$ in equation~\eqref{eq:switching_stress_tensor} take values between $0$ and $1$, it is found that there is not a smooth transition between the two extremes of behaviour. Instead, we find a critical value of $s^2$, between $0.5$ and $0.6$, below which (closer to isotropic) we see flows simple enough to create and sustain feedback reconnection, and above which (closer to anisotropic) we see flows complex enough to disrupt this feedback situation. This behaviour can be seen in how the kinetic energy time series changes with $s^2$ in Figure~\ref{fig:kinetic-energy-changing-s}.

\section{Parameter study}
\label{sec:results2}

In order to confirm that the results of Section~\ref{sec:results} are typical, and to further understand how they vary, two parameter studies are performed; one varying viscosity, keeping all other parameters constant; and one varying resistivity, again keeping all other parameters constant. 

In the first study we vary the viscosity as $\nu = 5 \times 10^{-n}$, where the index $n$ takes the values $4.75$, $4.5$, $4.25$, $4$ and $3.75$, while keeping resistivity constant at $\eta = 5\times10^{-4.5}$. This range of viscosities represents values that are typically used in simulations, with a lower bound above numerical diffusion and an upper bound below physically unrealistic values for the corona.

In the second study we vary the resistivity as $\eta = 5 \times 10^{-m}$, where the index $m$ takes the values $4.75$, $4.5$, $4.25$, $4$, $3.75$, and $3.5$, while keeping the resistivity constant at $\nu = 5\times 10^{-4.5}$. Similar to the limits on viscosity, any lower resistivities become comparable to numerical diffusion. Higher resistivities diffuse the field so quickly that the instability does not have time to grow.

\subsection{Effect on the secondary instability varying diffusion parameters}
\label{sec:secondary_instability}

\begin{figure}[t]
    \centering
    \begin{subfigure}[t]{0.5\textwidth}
      \centering
      \begin{overpic}[width=\textwidth]{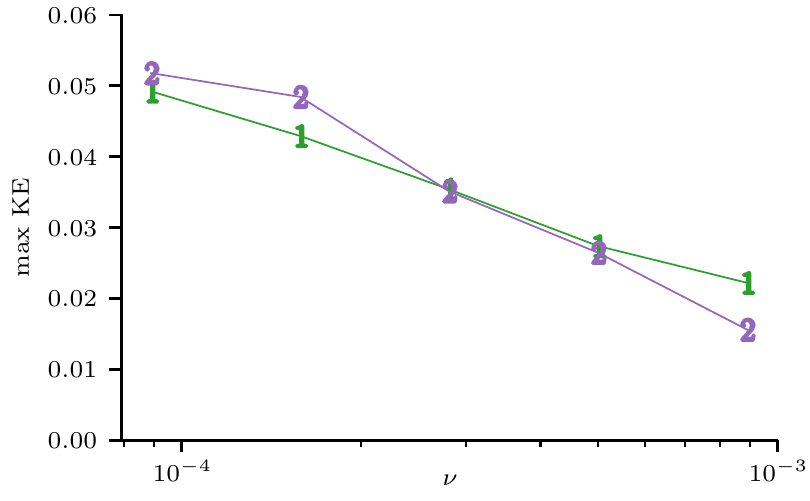}
        \put (50,62) {\small\textbf{(a)}}
      \end{overpic}
    \end{subfigure}%
    ~
    \begin{subfigure}[t]{0.5\textwidth}
      \centering
      \begin{overpic}[width=\textwidth]{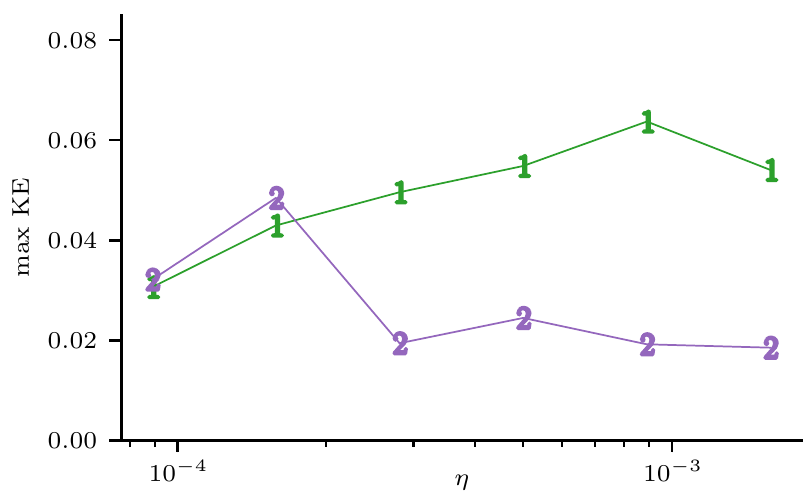}
        \put (50,62) {\small\textbf{(b)}}
      \end{overpic}
    \end{subfigure}
    \caption{\textit{Maximum kinetic energy corresponding to initial instability and secondary instability as functions of \rs{resistivity $\eta$ and viscosity $\nu$.}} \textbf{(a)} resistivity $\eta$ is fixed at $5\times10^{-4.5}$ and \textbf{(b)} viscosity $\nu$ is fixed at $5\times10^{-4.5}$. In both plots are shown the maximum kinetic energy produced by the initial instability (green, $1$-marker) and the maximum kinetic energy produced by the secondary instability (purple, $2$-marker). Only results using the isotropic viscosity are shown.}
    \label{fig:secondary_instability}
\end{figure}

\ro{Figure~\ref{fig:secondary_instability} shows the maximum kinetic energy produced by the two instabilities found in the isotropic case in Section~\ref{sec:results}. The maximum kinetic energy provides a useful measure of the efficacy of an instability, particularly when comparing the relative magnitudes of the initial and secondary instabilities. Since we only find evidence of the secondary instability in the isotropic case, we do not show the results from the switching case.}

\ro{Looking at Figure~\ref{fig:secondary_instability}(a), it is observed that increasing $\nu$ reduces the kinetic energy generated in both instabilities. For small values of $\nu$ we find the secondary instability causes more energy to be produced than the first, however as $\nu$ increases, this relationship reverses, with the initial instability causing more energy to be produced than the secondary one for large $\nu$. This reversal suggests that the greater kinetic energy produced by the initial instability for low values of $\nu$ is causing a stronger current sheet to form, enhancing reconnection, and producing a stronger secondary instability.}

\ro{The effect of resistivity $\eta$ on the secondary instability is to suppress it entirely when $\eta$ is large. Since the secondary instability is driven by reconnection outflows, it is not surprising that we can find values of $\eta$ for which the reconnection outflows do not feedback to produce the instability.}

\subsection{Varying viscosity}

\label{sec:visc_param_study}

\subsubsection{Dependence of heating on viscosity}

\begin{figure}[t]
    \centering
    \begin{subfigure}[t]{0.32\textwidth}
      \centering
      \begin{overpic}[width=\textwidth]{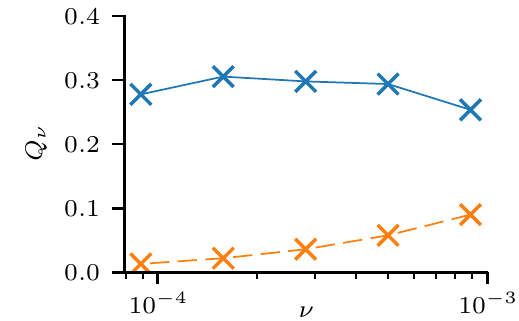}
        \put (50,60) {\small\textbf{(a)}}
      \end{overpic}
    \end{subfigure}%
    ~
    \begin{subfigure}[t]{0.32\textwidth}
      \centering
      \begin{overpic}[width=\textwidth]{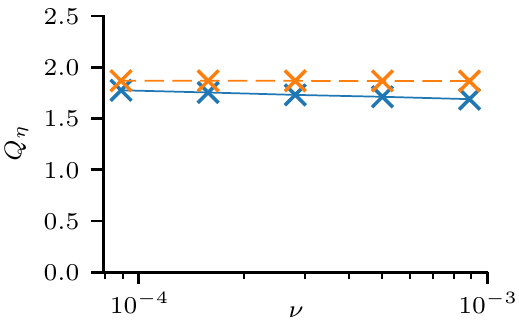}
        \put (50,60) {\small\textbf{(b)}}
      \end{overpic}
    \end{subfigure}
    ~
    \begin{subfigure}[t]{0.32\textwidth}
      \centering
      \begin{overpic}[width=\textwidth]{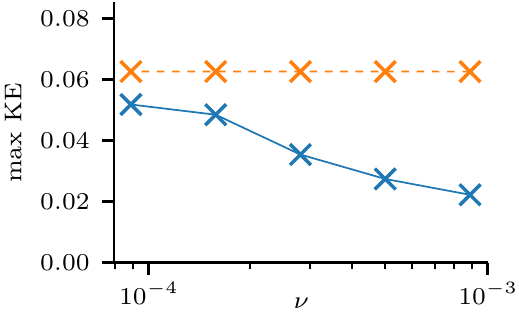}
        \put (50,60) {\small\textbf{(c)}}
      \end{overpic}
    \end{subfigure}
    \caption{\textit{Anisotropic viscous heating, Ohmic heating, and maximum kinetic energy as functions of \rs{viscosity $\nu$ for a fixed resistivity $\eta=5\times10^{-4.5}$.}} \textbf{(a)} Total viscous heat, \textbf{(b)} total Ohmic heat, and \textbf{(c)} maximum (in time) kinetic energy produced using isotropic viscosity (blue, solid) and switching viscosity (orange, dashed) as functions of viscosity $\nu$ at the final time of $t=400$. The anisotropic viscous heating has been multiplied by a factor of $10$. The maximum kinetic energy is calculated as the maximum value prior to $t=400$.}
    \label{fig:param_study_varying_viscosity}
\end{figure}

Figures~\ref{fig:param_study_varying_viscosity}(a) and~\ref{fig:param_study_varying_viscosity}(b) show \rr{the} total heat generated by $t=400$ via viscous $Q_{\nu}$ and Ohmic $Q_{\eta}$ dissipation as we vary the strength of viscosity $\nu$. \ro{It should be noted that, to allow the trend in the anisotropic viscous heating to be seen in the plot, it has been multiplied by a factor of $10$.} Before discussing the apparent trends in the heating when we vary viscosity, \rr{it is} useful to note that, \rt{just as in the typical case described previously, for the range of $\nu$ shown,} isotropic viscous heating remains approximately two orders of magnitude greater than the anisotropic viscous heating, and the Ohmic heating is consistently higher when using anisotropic viscosity than when using isotropic.

\ro{Since viscous dissipation (equations~\eqref{eq:iso_viscous_heating} and~\eqref{eq:aniso_viscous_heating}) has a functional dependence on $\nu$ and Ohmic dissipation (equation~\eqref{eq:ohmic_heating}) does not, we could naively assume that as we vary viscosity we should see some trend in the viscous dissipation for both models and no trend in the Ohmic dissipation. The trends that are observed broadly adhere to this but, unexpectedly we do find some trend in the Ohmic heating when using isotropic viscosity.}

\ro{When employing the switching model, the Ohmic heating appears to be independent of $\nu$, whereas when employing the isotropic model, we find a small trend of decreased Ohmic heating with increased $\nu$. These trends can be explained by considering the effect of viscosity on compressive flows and current densities. During the kink instability, Ohmic heating, being proportional to the square of the local current density, is increased when an already sheared magnetic field is compressed by flows perpendicular to the field, increasing the local current density. Thus, as the speeds of perpendicular flows increase, so does the Ohmic heating. These perpendicular flows are effectively only damped by isotropic viscosity. Since we find the maximum kinetic energy (Figure~\ref{fig:param_study_varying_viscosity}(c)) decreases with $\nu$ in \emph{only} the isotropic case, and remains constant in the switching case, it is appropriate that the Ohmic heating decreases with $\nu$ in the isotropic case and is negligibly dependent on $\nu$ in the switching case.}

\ro{If varying $\nu$ does not change the dynamics in the switching case, the functional dependence of $Q_{\nu}$ on $\nu$ (see equation~\eqref{eq:aniso_viscous_heating}) suggests we should observe an increase in anisotropic viscous heating with $\nu$. Figure~\ref{fig:param_study_varying_viscosity}(a) reveals precisely this.}

\ro{The relationship between the isotropic viscous heating at $\nu$ appears non-trivial. Given the decrease in maximum kinetic energy (Figure~\ref{fig:param_study_varying_viscosity}(a)) with $\nu$, it is expected the isotropic viscous heating should also decrease. However, this is not what is observed. Although there appears to be a slight decreasing trend in the isotropic viscous heating when $\nu$ is increased past $10^{-4}$, the left-most point is clearly an outlier. This suggests the secondary instability is having a significant and non-trivial effect on the heating. Indeed this is also suggested by the subtle change of gradient in the maximum kinetic energy on the left-hand side of the Figure~\ref{fig:param_study_varying_viscosity}(c). Due to this particular parameter study producing only five data points, we cannot discuss these trends with much confidence. A more detailed parameter study should be performed, investigating more values of $\nu$ within and beyond the range studied here.}

\subsubsection{Dependence of linear growth rate on viscosity}
\label{sec:linear_growth_rate_varying_visc}

\begin{figure}[t]
    \centering
    \begin{subfigure}[t]{0.5\textwidth}
      \centering
      \begin{overpic}[width=\textwidth]{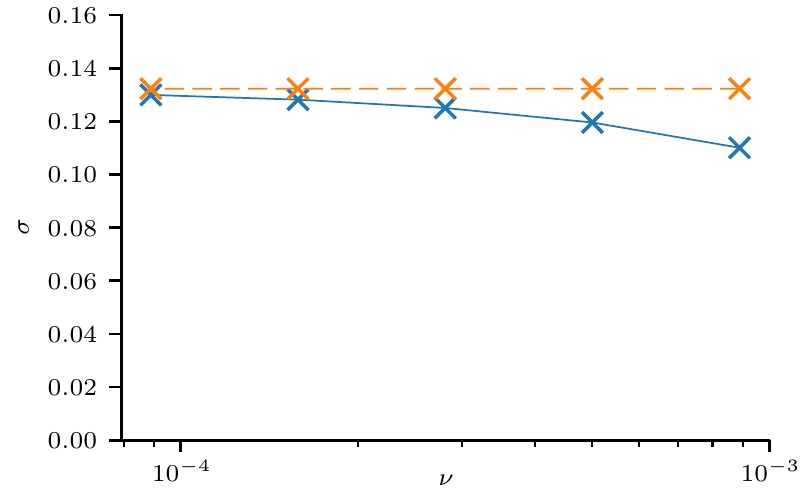}
        \put (50,62) {\small\textbf{(a)}}
      \end{overpic}
    \end{subfigure}%
    ~
    \begin{subfigure}[t]{0.5\textwidth}
      \centering
      \begin{overpic}[width=\textwidth]{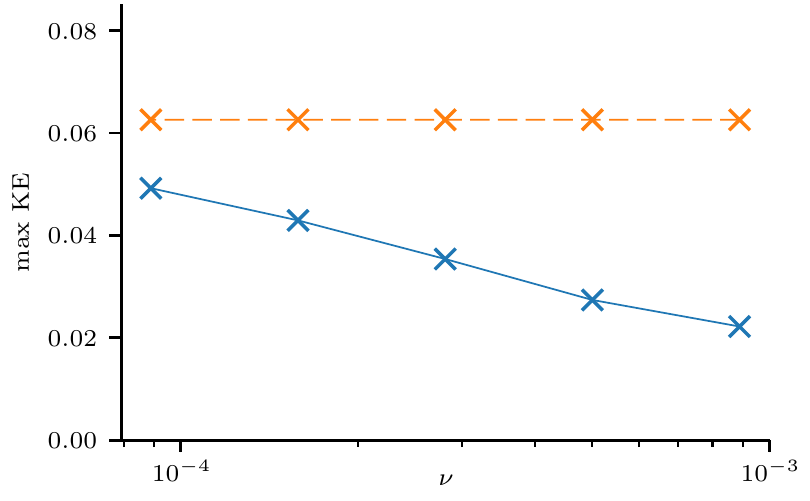}
        \put (50,62) {\small\textbf{(b)}}
      \end{overpic}
    \end{subfigure}
    \caption{\textit{Linear growth rate and maximum (in time) kinetic energy as
        functions of \rs{viscosity $\nu$ for a fixed
          resistivity of $\eta=5\times10^{-4.5}$.}} \textbf{(a)} Growth rate and \textbf{(b)} maximum kinetic energy, generated by isotropic viscosity (blue, solid) and switching viscosity (orange, dashed) as functions of visocity $\nu$. The maximum kinetic energies are calculated as the maximum values in time prior to $t=125$. This is to capture the behaviour of only the initial nonlinear evolution of the instability, neglecting any further instabilities like the secondary instability found in Section~\ref{sec:results}. Note, the maxima do not necessarily occur at the same time and this particular parameter study has been performed fixing $\nu$ at a slightly different value to the previous parameter studies.}
    \label{fig:growth_rate_varying_viscosity}
\end{figure}

\rt{For each value of $\eta$ we calculate the linear growth rate $\sigma$ of the onset of the kink instability by plotting the logarithm of the kinetic energy against time and measuring the gradient during the period of linear growth (as is done in Figure~\ref{fig:log_kinetic_energy_over_time}). Figure~\ref{fig:growth_rate_varying_viscosity}(a) plots these growth rates against $\eta$, and Figure~\ref{fig:growth_rate_varying_viscosity}(b) shows the maximum kinetic energy calculated as the maximum prior to $t=125$. For every $\eta$, this time is between the peaks of the kinetic energy corresponding to the first and secondary instabilities. Taking the maximum before this time allows us to capture only the behaviour of the initial instability, since this is the instability of interest in this section.}

\rt{It can be seen from the relationship between the growth rate and $\nu$ for both viscosity models that isotropic viscosity appears to begin to suppress the kink instability, for larger $\nu$, while the switching viscosity does not (Figure~\ref{fig:growth_rate_varying_viscosity}(a)). This is also apparent from the relationship between the maximum kinetic energy and $\nu$ for both models (Figure~\ref{fig:growth_rate_varying_viscosity}(b)). This difference between the viscosity models results from the anisotropic viscosity being so weak that the dynamics of the initial onset of the kink instability are not significantly affected by a significant increase in $\nu$.}

\subsection{Varying resistivity}

\subsubsection{Dependence of heating on resistivity}

\begin{figure}[t]
    \centering
    \begin{subfigure}[t]{0.5\textwidth}
      \centering
      \begin{overpic}[width=\textwidth]{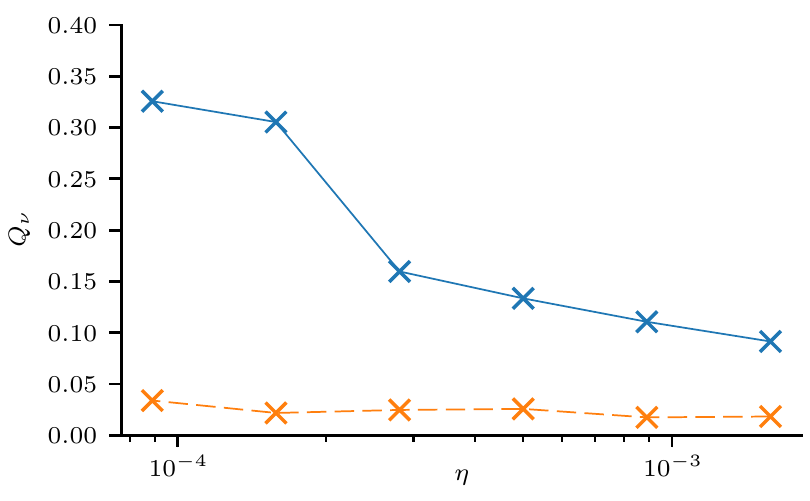}
        \put (50,60) {\small\textbf{(a)}}
      \end{overpic}
    \end{subfigure}%
    ~
    \begin{subfigure}[t]{0.5\textwidth}
      \centering
      \begin{overpic}[width=\textwidth]{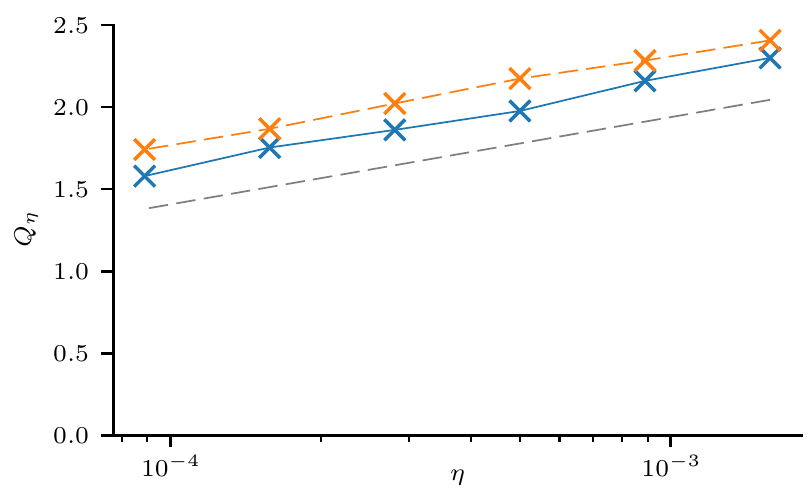}
        \put (50,60) {\small\textbf{(b)}}
      \end{overpic}
    \end{subfigure}
    \caption{\textit{Anisotropic viscous, and Ohmic heating as functions of \rs{resistivity $\eta$ for a fixed value of
          viscosity $\nu=5\times10^{-4.5}$.}} \textbf{(a)} Total viscous heat and \textbf{(b)} total Ohmic heat produced using generated by isotropic viscosity (blue, solid) and switching viscosity (orange, dashed) as functions of resistivity $\eta$ at the final time of $t=400$. The anisotropic viscous heating has been multiplied by a factor of $10$. \rt{Overlaid on figure (b) is the scaling $\log_{10}(\eta^{1/2})$.}}
    \label{fig:param_study_varying_resistivity}
\end{figure}

Figure~\ref{fig:param_study_varying_resistivity}(a) and~\ref{fig:param_study_varying_resistivity}(b) show the total heating generated by $t=400$ via viscous $Q_{\nu}$ and Ohmic $Q_{\eta}$ dissipation as we vary the strength of resistivity $\eta$. Just as in Section~\ref{sec:visc_param_study}, the anisotropic viscous heating is multiplied by $10$. Again, \rr{it is} useful to note that, across the entire range of $\eta$ studied here, the isotropic viscous heating remains approximately two orders of magnitude greater than the anisotropic viscous heating and the Ohmic heating produced when using switching viscosity is consistently higher than that produced when using isotropic viscosity. This aligns with the results when varying viscosity, as discussed above.

\rt{As in the parameter study varying $\nu$, we also observe a non-trivial relationship between the viscous heating for both models and $\eta$ (Figure~\ref{fig:param_study_varying_resistivity}(a)). The isotropic viscous heating reveals an decreasing trend over all values of $\eta$ studies here, however there is a clear jump in heating between approximately $10^{-3.9}$ and $10^{-3.7}$. Given that these are the values of $\eta$ where we observe strong influence of the secondary instability on the kinetic energy output (see Section~\ref{sec:secondary_instability}), these results suggest it is the kinetic energy produced by the secondary instability that is being damped at low values of $\eta$.}

\rt{Just as in the parameter study varying $\nu$, the anisotropic viscosity shows very little variability with $\eta$ (Figure~\ref{fig:param_study_varying_resistivity}(a)), even with the heating multiplied by a factor of $10$ for plotting. Despite the dynamics significantly changing with $\eta$, the effect of the anisotropic viscosity is so small that we observe very little change in the heating.}

\rt{We observe that Ohmic heating increases with increasing $\eta$ (Figure~\ref{fig:param_study_varying_resistivity}(b)). This is to be expected given the functional dependence of $Q_{\eta}$ on $\eta$, however the actual scaling is not linear in $\eta$, as might be predicted from equation~\eqref{eq:ohmic_heating}. Rather, we find that $Q_{\nu}$ varies linearly with $\log_{10}(\eta^{1/2})$ for the range of $\eta$ studied here. Without a more comprehensive parameter study covering more values of $\eta$, it is difficult to reason why the scaling takes this form. However, we are able to state with confidence that the use of anisotropic viscosity is consistently enhancing Ohmic heating across the range of $\eta$ studied here. This is due to the kink instability producing more kinetic energy in the switching case, which better compresses the magnetic field, creating stronger current sheets and thus enhancing Ohmic heating.}

\subsubsection{Dependence of linear growth rate on resistivity}

\begin{figure}[t]
    \centering
    \begin{subfigure}[t]{0.5\textwidth}
      \centering
      \begin{overpic}[width=\textwidth]{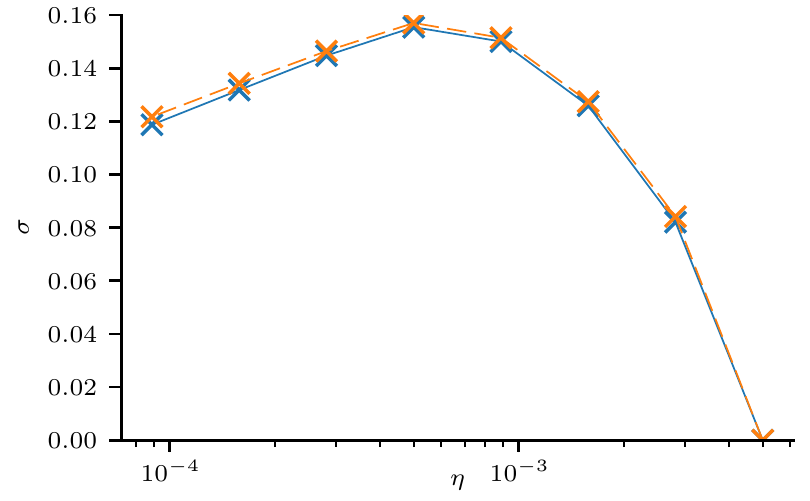}
        \put (50,62) {\small\textbf{(a)}}
      \end{overpic}
    \end{subfigure}%
    ~
    \begin{subfigure}[t]{0.5\textwidth}
      \centering
      \begin{overpic}[width=\textwidth]{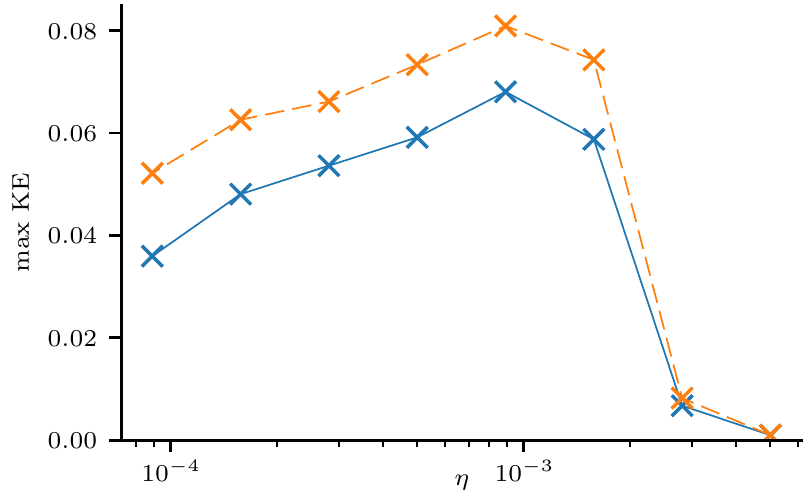}
        \put (50,62) {\small\textbf{(b)}}
      \end{overpic}
    \end{subfigure}
    \caption{\textit{Linear growth rate and maximum (in time) kinetic energy as
        functions of \rs{resistivity $\eta$ for a fixed
          viscosity of $\nu=10^{-4}$.}} \textbf{(a)} Growth rate and \textbf{(b)} maximum kinetic energy, generated by isotropic viscosity (blue, solid) and switching viscosity (orange, dashed) as functions of resistivity $\eta$. The maximum kinetic energies are calculated as the maximum values in time prior to $t=125$. This is to capture the behaviour of only the initial nonlinear evolution of the instability, neglecting any further instabilities like the secondary instability found in Section~\ref{sec:results}. Note, the maxima do not necessarily occur at the same time and this particular parameter study has been performed fixing $\nu$ at a slightly different value to the previous parameter studies.}
    \label{fig:growth_rate_varying_resistivity}
\end{figure}

\rt{As is done in Section~\ref{sec:linear_growth_rate_varying_visc}, we calculate the linear growth rates for each value of $\eta$. These and the maximum early time ($t<125$) kinetic energy are shown in Figure~\ref{fig:growth_rate_varying_resistivity}. The plots show that the use of the switching model seems to consistently amplify the growth of the kink instability, shown both in the growth rate and in the kinetic energy. Beyond this, the two models of viscosity show similar trends with $\eta$.}

\rt{Both plots in Figure \ref{fig:growth_rate_varying_resistivity} show that the kink instability is strongly inhibited for values of $\eta$ greater than approximately $10^{-2.5}$. This can be explained by the initial diffusion of the magnetic field being so fast-acting for large values of $\eta$ that the instability is totally suppressed. The increased suppression of the instability with strength of Ohmic diffusion can be seen in both plots as $\eta$ increases past $10^{-3}$.}

\section{Summary and discussion}
\label{sec:conclusions}

We have studied the linear and nonlinear development of the MHD kink
instability with two different viscosity models. The first is
isotropic (Newtonian) viscosity, which is the most commonly used
viscosity model \rs{in coronal loop studies}. The second is anisotropic viscosity, representing the strong-field limit of Bragkinskii viscosity with a preferred direction parallel to the magnetic field. The implementation of anisotropic viscosity is via the switching model~\cite{mactaggartBraginskiiMagnetohydrodynamicsArbitrary2017} which is suitable for coronal applications.

By considering particular (low) values of the viscosity and resistivity, we find that the effect of the different viscosity models on the linear onset of the kink instability is marginal. The significant differences appear in the nonlinear phase. Two main phases of evolution can be identified which highlight the differences between the effects of the two viscosity models. The anisotropic \rr{(switching)} case produces more kinetic energy at the onset of the nonlinear phase of the instability---the first phase. It also produces flows and current sheets with smaller length scales compared to the isotropic case and this allows the magnetic field to relax faster due to more efficient reconnection. 

In the second phase, the isotropic case exhibits a secondary instability, which is not found in the anisotropic case. This new instability leads to enhanced reconnection and faster magnetic relaxation, compared to the anisotropic case.  The simulations are run for $600$ Alfv\'en times (a long time period for coronal applications) and the behaviour of the second phase continues for all of this time.

We have also run a series of parameter studies varying the strengths of viscosity and resistivity. We find the qualitative results of the two phases of the detailed investigation hold true over a range of viscosities and resistivities, including the existence of the secondary instability. Notably, over all parameters studied, viscous heating is consistently overestimated by the isotropic model, and Ohmic heating is consistently enhanced by use of the switching model.

Although there can be much variability in the nonlinear behaviour of the kink instability, our results reveal an important general finding. At the beginning of the nonlinear phase, anisotropic (parallel) viscosity allows for the development of smaller length scales (both flows and current sheets), \rr{compared to isotropic viscosity,} leading to more efficient reconnection and faster magnetic relaxation, at least initially. As has been described in detail, isotropic viscosity can produce other effects later. However, for coronal applications, it is the initial nonlinear phase of the instability that is likely to be of most interest since, in reality, a coronal loop will interact with others on a longer time scale, \rr{thus affecting the nonlinear evolution}. Since anisotropic viscosity is a more realistic model for viscosity in the corona, our results will be useful in the interpretation of observations of coronal loops that are kink unstable. This topic will be an interesting avenue of future research.

\section*{Acknowledgements}

Results were obtained using the ARCHIE-WeSt High Performance Computer
(\url{www.archie-west.ac.uk}) based at the University of
Strathclyde. JQ was funded via an DTA EPSRC studentship. 

\section*{Associated software}

The specific version of lare3d used can be found via Zenodo~\cite{tonyarber_2019_3560251}. The data analysis and instructions for reproducing all results found in this report may be also found at~\cite{jamie_j_quinn_2019_3560245}. Finally, our field line integration tool may be found at~\cite{jamie_j_quinn_2019_3560249}.

\bibliography{paper}

\end{document}